\documentclass{aa} 

\usepackage{graphicx}
\usepackage{mathtools} 
\usepackage[varg]{txfonts} 
\usepackage{xcolor}
\usepackage{xspace}
\usepackage{multirow}
\usepackage{rotating, lscape}
\usepackage{float}

\usepackage{hyperref}

\graphicspath{{./}{figs/}}

\def\J{\textit{J}\xspace}
\def\H{\textit{H}\xspace}
\def\Ks{\textit{$K_s$}\xspace}

\def\Av{$A_{\rm V}$\xspace}

\def\NHth{NH$_3$\xspace}
\def\NtHp{N$_2$H$^+$\xspace}
\def\NtDp{N$_2$D$^+$\xspace}
\def\DCOp{DCO$^+$\xspace}
\def\Hp{H$^+$\xspace}
\def\Ht{H$_2$\xspace}

\def\Hthp{H$_3^+$\xspace}
\def\HtDp{H$_2$D$^+$\xspace}

\def\oHtDpgrd{1$_{10}$--1$_{11}$\xspace}
\def\DtHp{D$_2$H$^+$\xspace}
\def\Dthp{D$_3^+$\xspace}
\def\CetO{C$^{18}$O\xspace}
\def\thtCO{$^{13}$CO\xspace}
\def\twvCO{$^{12}$CO\xspace}

\def\arcmin{\mbox{$^{\prime}$}\xspace}
\def\arcsec{\mbox{$^{\prime\prime}$}\xspace}
\newcommand\micron{\mbox{$\mu$m}\xspace}

\begin{document}

\title{Physical and chemical modeling of the starless core \object{L\,1512}
       \thanks{Based in part on observations obtained with WIRCam, a joint project of CFHT, Taiwan, Korea, Canada, France, at the Canada-France-Hawaii Telescope (CFHT), which is operated by the National Research Council (NRC) of Canada, the Institute National des Sciences de l'Univers of the Centre National de la Recherche Scientifique of France, and the University of Hawaii. The observations at the Canada-France-Hawaii Telescope were performed with care and respect from the summit of Maunakea which is a significant cultural and historic site.}
       }

\author{Sheng-Jun Lin\inst{1}
        \and Laurent Pagani\inst{2}
        \and Shih-Ping Lai\inst{1}
        \and Charl\`ene Lef\`evre\inst{3}
        \and Fran\c{c}ois Lique\inst{4} 
        }

\institute{Institute of Astronomy, National Tsing Hua University, No.\,101, Section 2, Kuang-Fu Road, Hsinchu 30013, Taiwan\\
           \email{sj.lin@gapp.nthu.edu.tw, slai@phys.nthu.edu.tw}
           \and
           LERMA \& UMR8112 du CNRS, Observatoire de Paris, PSL  University, Sorbonne Universit\'es, CNRS, F-75014 Paris, France
           \and
           Institut de Radioastronomie Millim\'etrique (IRAM), 300 rue de la Piscine, 38400 Saint-Martin d’H\`eres, France
           \and
           LOMC - UMR 6294, CNRS-Universit\'e du Havre, 25 rue Philippe Lebon, BP 1123, 76063, Le Havre, France
           }

\date{Received abc, dd, yyyy; accepted abc, dd, yyyy}

\abstract{The deuterium fractionation in starless cores gives us a
clue to estimate their lifetime scales, thus allowing us to
distinguish between dynamical theories of core formation.
Cores also seem to be subject to a differential N$_2$ and CO
depletion, which was not expected from the models.}
{We aim to create a survey of ten cores to estimate their lifetime scales and depletion profiles in detail. After describing L\,183, located in Serpens,
we present the second cloud of the series, L\,1512, from the star-forming region Auriga.}
{To constrain the lifetime scale, we performed chemical modeling of the deuteration profiles across L\,1512 based on dust extinction
measurements from near-infrared observations and nonlocal thermal equilibrium radiative transfer with multiple line observations
of \NtHp, \NtDp, \DCOp, \CetO, and $^{13}$CO, plus \HtDp (\oHtDpgrd).}
{We find a peak density of 1.1$\times$10$^5$ cm$^{-3}$ and a central temperature of 7.5$\pm$1 K, which are higher and
lower, respectively, compared with previous dust emission studies. The depletion factors of \NtHp and \NtDp are 27$^{+17}_{-13}$ and
4$^{+2}_{-1}$ in L\,1512, which are intermediate between the two other more advanced and denser starless core cases, L\,183 and L\,1544.
These factors also indicate a similar freeze-out of N$_2$ in L\,1512, compared to the two others despite a peak density one to two
orders of magnitude lower.
Retrieving CO and N$_2$ abundance profiles with the chemical model,
we find that CO has a depletion factor of $\sim$430--870 and the N$_2$ profile is similar to that of CO unlike that toward L\,183.
Therefore, L\,1512 has probably
been living long enough so that N$_2$ chemistry has reached steady state.}
{\NtHp modeling is necessary to assess the precise physical conditions in the center of cold starless cores, rather than
dust emission. L\,1512 is presumably older than 1.4 Myr. Therefore, the dominating core formation mechanism should be
ambipolar diffusion for this source.}


\maketitle

\section{Introduction}

Starless cores are the sites of future star and planet formation. Although we know gravitational collapse plays a main role
during star formation, the details of the collapse process, particularly the timescale, are not yet well understood. There are
two major dynamical theories with different dominating mechanisms. One theory is the ambipolar diffusion-controlled scenario, advocating
that molecular clouds evolve quasi-statically to form starless cores after 1--10 Myr \citep{Shu87, Tassis04, Mouschovias06,
McKee07}. The other theory is the supersonic turbulence-induced scenario, in which cores can form faster through gravo-turbulent
fragmentation with a free-fall timescale less than 1 Myr \citep{MacLow98, Padoan99, Hartmann01, Federrath12}. Measuring the age of the cores would
therefore allow us to select the correct scenario but the task is difficult. Time-dependent chemical analysis has long been
thought to be a solution to this problem but unknown initial conditions too often hamper a precise determination of the elapsed
time from the molecular cloud assembly to the formation of the core (see, e.g., \citealt{Nilsson00} for a discussion on the
sensitivity of a chemical model to initial conditions). To circumvent the initial condition problem, \citet{Pagani09b,Pagani13}
proposed an approach based on the \NtDp/\NtHp deuteration profile across the starless core. These works showed that the profile
itself is mostly dependent on the time it took to form the starless core, showing a clear distinction between fast collapsing
cores with steep profiles and slow collapsing cores with much higher deuteration level and much flatter profiles. This is possible
thanks to the slow decrease of the ortho/para-\Ht (thought to be originally produced on grains in the 3:1 ratio), which is
comparable to the lifetime of starless cores and because a high abundance of ortho-\Ht prevents the deuteration to proceed
\citep{Pagani92b,Pagani09b,Pagani11}. Varying the initial conditions did not change the clear-cut separation between the two types
of profiles. L\,183 was shown to clearly belong to the fast collapse timescale (in a free-fall time; \citealt{Pagani13}). More
recently, other dynamical studies also based on deuteration (but not on \NtHp deuteration radial profile evolution) have been
published \citep{Kong15,Lackington16,Kortgen17,Kortgen18,Giannetti19}; but these works have contradictory
conclusions concerning the typical lifetime of cores.

\begin{table*}
\centering
    \caption{Observational parameters.}
    \begin{tabular}{lcrccccc}
    \hline\hline
    Species & Transition & Frequency$^{\tablefootmark{a}}$ & $\delta$v & Beam Size$^{\tablefootmark{b}}$  & $\eta_{\rm MB}$ & $T_{\rm sys}$ $^{\tablefootmark{c}}$ & rms 
    noise\\ 
    & & (MHz) & (m\,s$^{-1}$) & (\arcsec) & & (K) & (mK)\\
    \hline
    \noalign{\smallskip}
    \multicolumn{8}{c}{IRAM 30-m (Dec 2013, May 2014, Oct 2014, Sep 2017)}\\
    \noalign{\smallskip}
    \hline
    \noalign{\smallskip}
    \NtHp & 1--0 & 93173.764 & 31 & 26 & 0.85 & 90--160 & 32--71\\
    \DCOp & 2--1 & 144077.282 & 41 & 17 & 0.80 & 80--300 & 28--102\\
    \NtDp & 2--1 & 154217.181 & 38 & 16 & 0.78 & 90--600 & 18--33\\
    \DCOp & 3--2 & 216112.582 & 68 & 11 & 0.66 & 360--460 & 53--94\\
    \CetO & 2--1 & 219560.358 & 67 & 11 & 0.65 & 340--400 & 50--89\\
    \NtHp & 3--2 & 279511.832 & 42 & 9 & 0.56 & 190--400 & 29--51\\
    \noalign{\smallskip}
    \hline
    \noalign{\smallskip}
    \multicolumn{8}{c}{GBT 100-m (Nov 2014)}\\
    \noalign{\smallskip}
    \hline
    \noalign{\smallskip}
    \DCOp & 1--0 & 72039.303 & 48 & 11 & 0.50 & 140--160 & 40--90\\
    \NtDp & 1--0 & 77109.616 & 44 & 10 & 0.50 & 105--130 & 29--44\\
    \noalign{\smallskip}
    \hline
    \noalign{\smallskip}
    \multicolumn{8}{c}{JCMT 15-m (Dec 2015)}\\
    \noalign{\smallskip}
    \hline
    \noalign{\smallskip}
    \HtDp & \oHtDpgrd & 372421.356 & 49 & 13& 0.70 & 690 & 19--71\\
    \NtHp & 4--3 & 372672.526 & 49 & 13 & 0.70 & 725 & 26--50\\
    \noalign{\smallskip}
    \hline
    \end{tabular}
    \tablefoot{
    \tablefoottext{a}{\NtHp and \NtDp frequencies are taken from 
    \citet{Pagani09a}. The frequencies correspond to the strongest
    hyperfine component for each transition.}
    \tablefoottext{b}{The beam size is the HPBW.}
    \tablefoottext{c}{$T_{\rm sys}$ is expressed in the $T_{\rm A}^*$ scale \citep{Kutner81}.}
    }
\label{tab:obs}
\end{table*}

The depletion effect on heavy species (e.g., CO, CS, SO) makes these species 
ineffective tracers of $n_{\rm H_2}$ and $T_{\rm kin}$ in starless cores.
Undepleted high-density tracers with low excitation energies are needed to reveal the inner structure of starless cores.
\NHth and \NtHp line emissions are commonly found to be spatially anti-correlated with
CO line emission among low- and high-mass starless cores,
which suggests that light nitrogen-bearing species are better tracers of the inner structure of starless cores \citep[e.g.,][]{Tafalla02, Fontani06, Pagani07}.
The \NtHp ($J$=1--0) rotational line 
provides a better constraint on both $n_{\rm H_2}$ and $T_{\rm kin}$ in dense cores than the \NHth ($J,K$)=(1,1) inversion line,
since the former has
a higher critical density ($5\times10^5$\,cm$^{-3}$) 
compared with the low critical density ($2\times10^3$\,cm$^{-3}$) of the latter \citep{Pagani07}.
However, when density is higher than $\sim$10$^6$ cm$^{-3}$,
\NtHp and its isotopologs start to decrease in abundance,
which suggests that their parent molecule, N$_2$, has frozen out onto dust \citep[e.g.,][]{Bergin02, Pagani05,Pagani07, Pagani12}.
In this case,  
since the CO- (and N$_2$-) depleted environment enhances the deuterium fractionation,
\Hp, \Hthp, and their isotopologs would be the most abundant ions.
Then \HtDp becomes the best tracer for the innermost region in starless cores
owing to the low excitation energy ($\Delta E/k=18$ K) of
the ortho-\HtDp ground transition, $J_{K_a K_c}$=\oHtDpgrd 
\citep{Khersonkii87, Pagani92a, Bergin02, VanDerTak05, Caselli08, Pagani09a}.

Many studies have estimated the depletion level in starless cores from column density maps, which imply a uniform density along the
line of sight \citep[e.g.,][]{Caselli99, Bacmann02, Bergin02, Kong16}. However, the depletion level at the center of cores could
be underestimated because of line-of-sight integration of their undepleted outer regions. \citet{Pagani07} performed spherically
symmetric ``onion shell''  modeling of \NtHp data toward L\,183, and showed a volume depletion of \NtHp by a factor of
6$_{-3}^{+13}$, which is larger than the depletion factor of $<$2 toward the same source estimated from the column density in
their earlier work \citep{Pagani05}. Similarly, \citet{Pagani12} show that CO depletion reaches a factor of 2000 instead of the
lower limit of 43 they found in an earlier work \citep{Pagani05}.

With the tools we developed \citep{Pagani07,Pagani09b,Pagani12,Pagani13} we can now expand our work, solely based on L\,183
until now, to other clouds. We have selected a list of ten clouds spread among several star-forming regions (Auriga, Taurus,
Serpens, and Rho Ophiuchus) for which we would like to study the dust content and dust growth (via the coreshine effect,
\citealt{Pagani10a,Steinacker10}), the CO and N$_2$ depletion profiles and the \NtDp/\NtHp deuteration profile. Dust growth,
depletion profile, and deuteration profile are three relatively independent ways to measure the age of starless cores and
therefore reinforce our confidence in the age determination of each of these cores. This allows us to address the discrepancies revealed
by the various works on dynamical deuteration mentioned above. Each core is studied in detail. From
the comparison of all the cores, we gain a better understanding of starless core formation and beyond, of low-mass star
formation. This first paper focuses on the gas phase of L\,1512.

L\,1512 is a relatively isolated nearby starless core that is located at a distance of 140 pc near the edge of the Taurus--Auriga molecular cloud complex \citep{Myers83a, Launhardt13}.
At 1 pc scale, L\,1512 is surrounded by CO gas elongated in the north-south direction with a sharp edge in the southern region \citep{Falgarone98, Falgarone01}.
The L\,1512 envelope is found to have no significant infall but probably oscillatory (or expansion) motion \citep{Lee01, Sohn07, Lee11}.
\citet{Kim16} further propose that the southern sharp edge is caused by external radiation from an A-type main-sequence star located at 97 pc away from us; this 
means a distance of $\sim$40 pc between the two if L\,1512 is at the Taurus--Auriga distance, which is a separation too large for a significant impact.
The L\,1512 core has a globular shape with a size of $\sim$0.08 pc, and is quiescent with line widths of $\sim$0.2 km s$^{-1}$ \citep{Cox89, Fuller93, Caselli95, Caselli02}.
The comparison between the HC$_3$N and \NHth line widths shows
a small nonthermal velocity dispersion of 0.038 km s$^{-1}$
and an averaged gas kinetic temperature of 11.6 K \citep{Fuller93}.
\NtHp observations reveal a nondepleted morphology at 0.1 pc scale,
but is found to be depleted in the innermost region of L\,1512
with an offset between the \NtHp and dust emission peaks \citep{Lee03}.

In this paper, we describe our observations in Sect.\,\ref{sec:observation} and present them in Sect.\,\ref{sec:results}.
In Sect.\,\ref{sec:analysis}, we perform the 1D spherically symmetric nonlocal thermal equilibrium (non-LTE) radiative transfer model and time-dependent chemical model.
In Sect.\,\ref{sec:discussion}, we discuss the lifetime and the possible formation mechanism of L\,1512. 
We summarize our results in Sect.\,\ref{sec:conclusion}.

\begin{figure}
	\centering
    \includegraphics[scale=0.3]{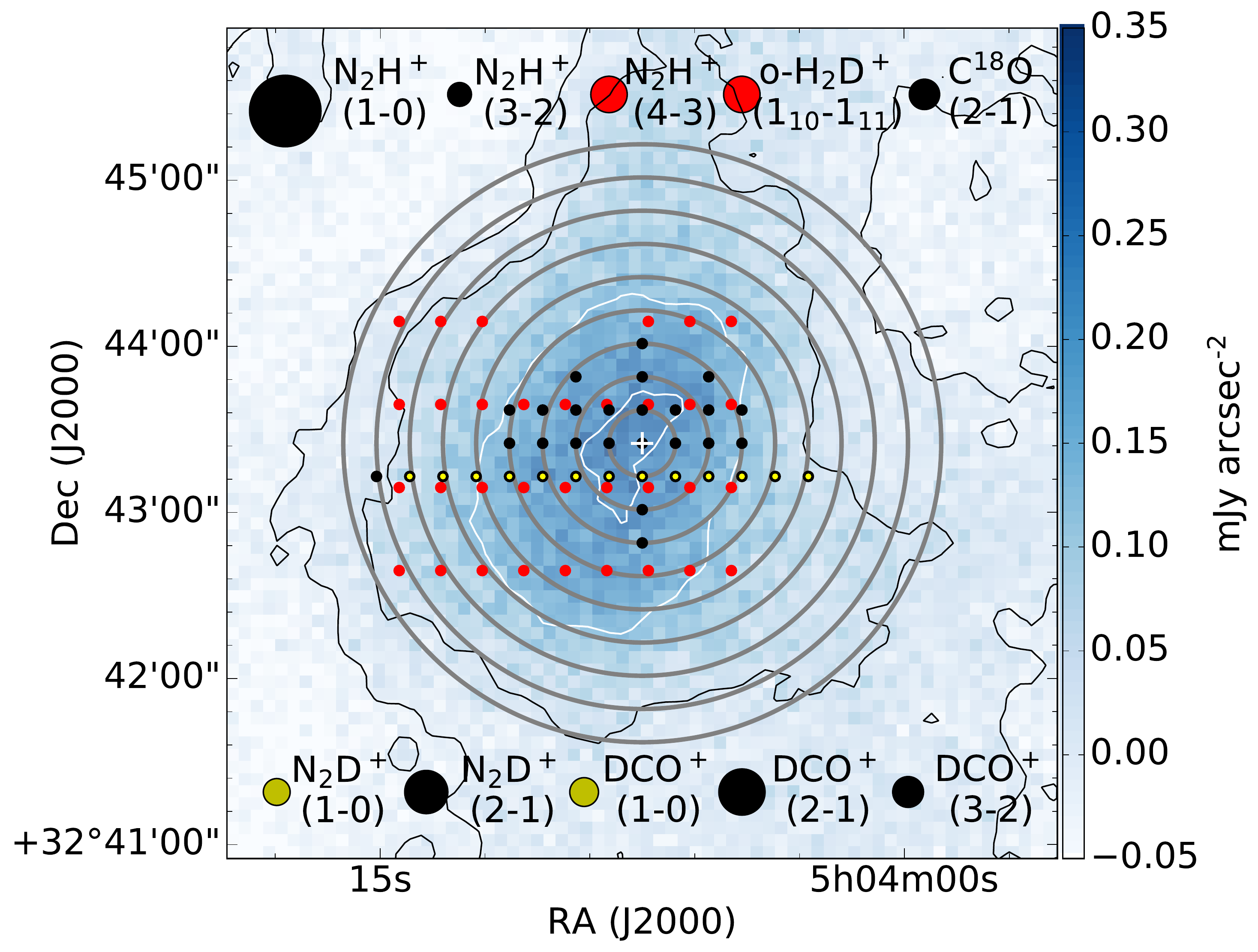}
	\caption{
    Multipointing grids and onion-shell model overlaid with the 850 \micron\,contour.
    The black and yellow dots in a ($\Delta$RA, $\Delta$Dec)=(12\arcsec, 12\arcsec)-spacing grid show the pointings of IRAM 30-m and GBT observations, respectively.
    The red dots in a ($\Delta$RA, $\Delta$Dec)=(15\arcsec,  30\arcsec)-spacing grid show those of JCMT HARP observations.
    The circles at top and bottom indicate the beam sizes of each spectral observations.
    The gray concentric circles with center of L1512 represent the onion-shell model with shell widths of 12\arcsec.
    The 850 \micron\,map is shown in blue-scale, and its white and black contours have the same levels as in Fig.\,\ref{fig:maps}g.}
    \label{fig:pointing}
\end{figure}

\section{Observations and data reduction}\label{sec:observation}

Observational parameters are summarized in Table\,\ref{tab:obs}
and the pointings of spectral line observations are shown in Fig.\,\ref{fig:pointing}.

\subsection{IRAM 30 m observations}

We observed L\,1512 using the IRAM 30 m telescope in December 2013, May and October 2014, and
September 2017.
Five molecular lines were observed in frequency-switching mode 
using the dual polarization Eight MIxer Receiver (EMIR), the Versatile Spectral Autocorrelator
(VESPA), and the Fourier Transform Spectrometer (FTS); 
these lines are \NtHp (1--0), \NtHp (3--2), \NtDp (2--1), \DCOp (2--1), and \DCOp (3--2).
The frequency resolution was adapted to keep the velocity sampling close to 50 m\,s$^{-1}$. Both
polarizations are systematically averaged.
The spectra were observed on a 12$\arcsec$ grid in an east-west strip 12\arcsec\ south of the core
center (RA, Dec)$_{\rm J2000}$ = 
(5$^{h}$04$^{m}$07\fs5, +32\degr43\arcmin25\farcs0) and along a north-south strip across
the core center. Pointing was regularly checked every 1.5 hours and pointing error was monitored to be less than 3\arcsec. 
Data were subsequently folded and baseline subtracted with CLASS\footnote{http://www.iram.fr/IRAMFR/GILDAS}. The original cut passes 12\arcsec\ south of
the true center because when the project started the CFHT and SCUBA-II maps were not yet available
and the core center position was not accurately known. Complementary observations in September 2017
(Director Discretionary Time) were performed. 
Two short east-west strips across the core center and 12\arcsec\ above the center, and one short 
strip along the north-south strip (mentioned above)
for \NtHp (1--0) and \DCOp (2--1)
allowed us to check that the modeling of the core is still consistent with the included offset.
Figure \ref{fig:pointing} shows the pointings of spectral line observations, and
the observational parameters are summarized in Table\,\ref{tab:obs}.
We complemented our observations with on-the-fly (OTF) maps of \NtHp (1--0) and \CetO (2--1)
obtained by \citet{Lippok13}, and multipointing mosaic maps of \CetO (1--0) and \thtCO (1--0) obtained by \citet{Falgarone98}.
\citet{Lippok13} and \citet{Falgarone98} give observational details.
The above data are regridded to supplement two orthogonal strips (Figs.\,\ref{fig:maincut_spectra} and \ref{fig:vercut_spectra})
and measure the spectral center velocity (Fig.\,\ref{fig:velocity structure}d) using a Gaussian fit routine in CLASS.
We also produce the moment maps of \NtHp (1--0) and \CetO (1--0) (Figs.\,\ref{fig:maps}h,i and Figs.\,\ref{fig:velocity structure}a,b). Among these, the \NtHp moment maps include our observations and the OTF data.

\subsection{JCMT observations}

The \NtHp (4--3) and \HtDp (\oHtDpgrd) observations were carried out in December 2015 using the JCMT 15 m telescope 
and the 16 pixels HARP receiver (two pixels of which are non-functioning) in  
frequency--switching mode. Thanks to the proximity of the two lines (250 MHz), they are routinely observed together with two sub-bands from the Auto 
Correlation Spectral Imaging System (ACSIS) spectrometer. Data are converted to the CLASS format to be reduced with CLASS (folding and baseline 
subtraction). Observational parameters are summarized in Table\,\ref{tab:obs},
and the pointings of spectral line observations are shown in Fig.\,\ref{fig:pointing}.

SCUBA-II observations were retrieved from the JCMT archive\footnote{http://www.cadc-ccda.hia-iha.nrc-cnrc.gc.ca/en/jcmt/}. They are part of the 
M13BC01 project (PI, James di Francesco). As far as we know, these data have not been published previously.

\subsection{GBT observations}

We observed L\,1512 using the GBT 100 m off-axis telescope in November 2014. We used the MM1 (67--74 GHz) and MM2 (73--80 GHz) W-band dual 
polarization sub-band receivers in in-band frequency-switching mode with the Versatile GBT Astronomical Spectrometer (VEGAS) backend. Pointing was 
checked every hour and each session started with a surface shape verification by using Out Of Focus Holography (OOF) on a strong source. Data were 
preprocessed (calibrated, folded, and baseline subtracted) in the GBTIDL data reduction program and converted to CLASS format for subsequent 
reduction. Thanks to the very smooth baseline of the receiver (due to the use of large band amplifiers instead of mixers), the preprocessing of the
data allows us to treat separately the ON and OFF observations (i.e., total power mode) and gain a factor of $\sqrt{2}$ on the noise (the method will be 
presented in a subsequent paper). 

\subsection{CFHT observations}
The CFHT Wide InfraRed CAMera (WIRCAM) was used with the wide filters \J, \H, and \Ks to observe the source on the night of 26 December 2013. To 
preserve the extended emission (scattered light known as cloudshine; \citealt{Foster06}), we used a SKY-TARGET-SKY nodding mode to subtract the atmospheric 
infrared emission. Seeing conditions were typically 0\farcs8. Data reduction was performed at the (now closed) TERAPIX center using a specific reduction method 
to recover the extended emission. The integration times were typically of 0.5 to 1 hour on-source per filter, to reach a completude magnitude of 21.5 (\J band) to 20 (\Ks band). The 
extended emission will be analyzed in a subsequent paper.

\begin{figure*}
	\centering
	\includegraphics[width=\textwidth]{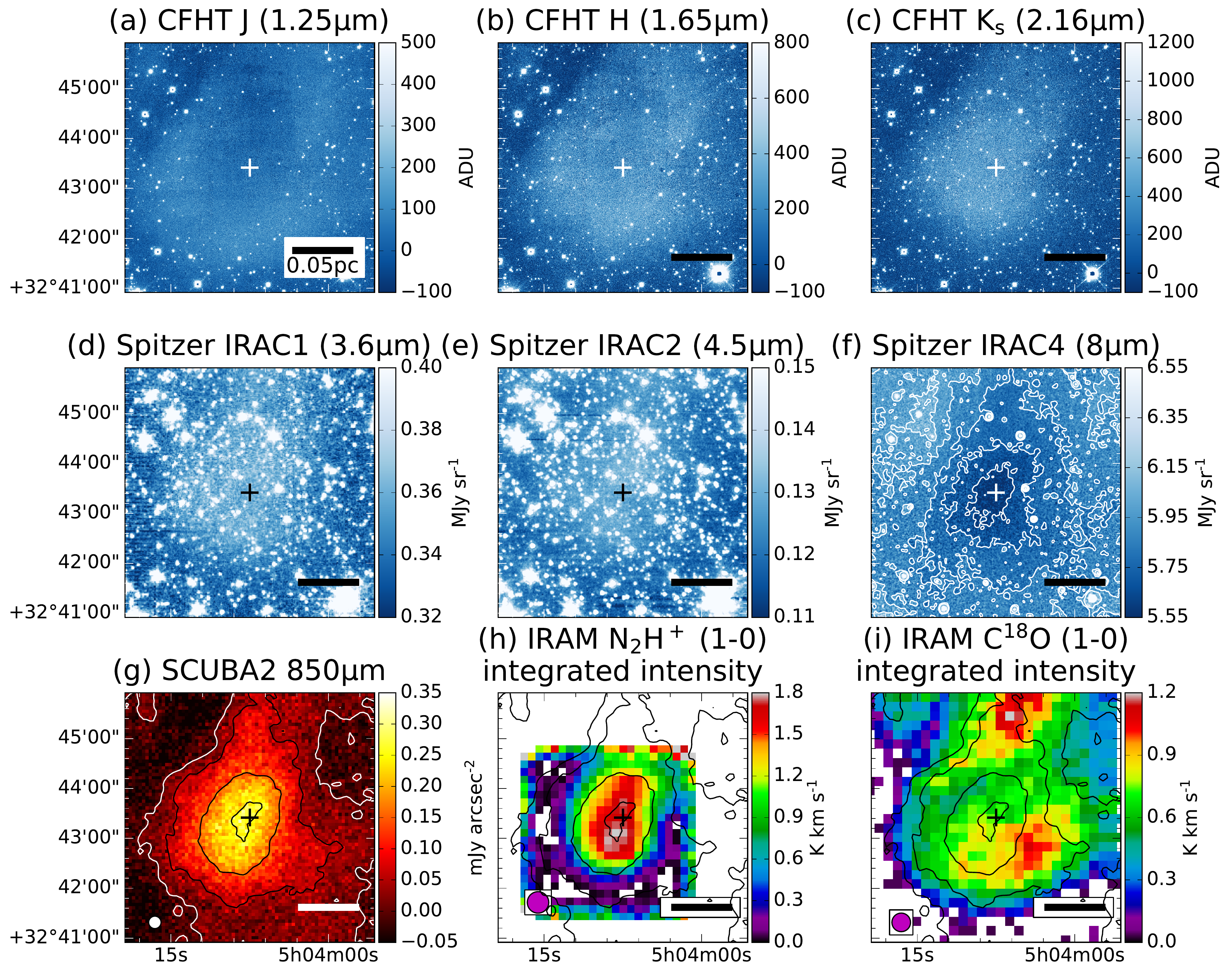}
	\caption{
    L1512 maps of continuum and line integrated intensity.
    The CFHT NIR maps at (a) \J band, (b) \H band, and (c) \Ks band.
    Spitzer MIR maps at (d) IRAC1 band, (e) IRAC2 band, and (f) IRAC4 band with its contours at 5.15, 5.25, 5.35, 5.45, 5.55 MJy/sr.
    (g) JCMT SCUBA map at 850 $\mu$m with its contours at 0, 0.05, 0.15, 0.25 mJy/arcsec$^2$.
    Integrated intensity maps of (h) \NtHp $J$=1--0 and (i) \CetO $J$=1--0 calculated within $V_{\rm LSR}$=[$-$3 km s$^{-1}$, 16 km s$^{-1}$] and [6.0 km s$^{-1}$, 7.5 km s$^{-1}$], 
    respectively.
    The panels $h$ and $i$ are overlaid with contours from the the 850 \micron\,image.
    The central cross in each panel indicates the center of L1512 determined by the IRAC4 map.
    The scale bars of 0.05 pc and millimeter-wavelength beam sizes 
    are denoted in the bottom right and bottom left corners, respectively.}
    \label{fig:maps}
\end{figure*}

\subsection{Spitzer observations} Spitzer observations are collected from the Spitzer Heritage Archive
(SHA)\footnote{https://sha.ipac.caltech.edu/applications/Spitzer/SHA/}. L\,1512 was observed in two programs with Spitzer InfraRed
Array Camera (IRAC), Program id. 94 (PI, Charles Lawrence) and Program id. 90109 (PI, Roberta Paladini); the second program occurred during the
warm period of the mission (only 3.6 and 4.5 $\mu$m channels still working). The P94 data were already discussed in
\citet{Stutz09}, while the warm mission data are analyzed in \citet{Steinacker15}. These two papers provide the
presentation of the observations. During the warm mission, deep observations of the cloud were performed and a completed magnitude
of 18 in band I2 (4.5\,\micron) was reached.

\section{Results}\label{sec:results}

\subsection{Continuum maps}

Figure\,\ref{fig:maps} shows the continuum maps of L1512 at near-infrared (NIR), mid-infrared (MIR), and submillimeter (submm) wavelengths. 
The NIR maps of L1512 at \J, \H, and \Ks bands (Figs.\,\ref{fig:maps}a--c) reveal extended emission that is distributed from an annular shape at \J band and progressively merges 
to a concentrated shape at \Ks band.
These structures are caused by the scattered light of ambient interstellar radiation field by dust grains ($a_{\rm gr}\sim0.1$ \micron) at the periphery of L1512.
Similar structures in dark clouds in the Perseus molecular complex have been found and studied by \citet{Foster06} and \citet{Juvela06}.
\citet{Foster06} introduced this scattering effect as cloudshine, and proposed to use it as a general tracer of the column density of dense clouds.
Since the wavelength of the maximum scattering cross section increases with the size of dust grains, the different spatial spread of the cloudshine of L1512 at the three bands also indicates that the dust grains have the smallest size at the outskirts, and become larger toward the inner region as discussed in \citet{Steinacker10} and \citet{Lefevre14, Lefevre16}.

On the other hand, the MIR maps of L1512 from IRAC1 and IRAC2 bands (Figs.\,\ref{fig:maps}d,e) show a more compact emission feature, while the IRAC4 map 
(Fig.\,\ref{fig:maps}f) shows  absorption instead.
The absorption feature is mainly caused by the bright PAH emission from the diffuse medium, the extinction of which becomes superior 
to the inner scattered light contribution \citep{Lefevre16}.
The similar spatial distributions between the compact emission and absorption indicate that the emission also comes from the central core.
\cite{Pagani10a} and \cite{Steinacker10} firstly recognized this phenomenon in their sample of nearby starless and Class 0/I protostellar cores.
By analogy to cloudshine, they named this effect coreshine.
With the benefit of the smaller beam sizes of IRAC maps compared with those of JCMT SCUBA maps,
we determined the center of L1512 core at (RA, Dec)$_{\rm J2000}$ = (5$^{h}$04$^{m}$07\fs5, +32\degr43\arcmin25\farcs0) 
using IRAC4 band (8 \micron) with a beam size of 2\arcsec,
and denote it as a cross shown in each panel of Fig.\,\ref{fig:maps}.

\begin{figure*}
\vspace{-0.3cm}
	\centering
	\includegraphics[height=23.5cm]{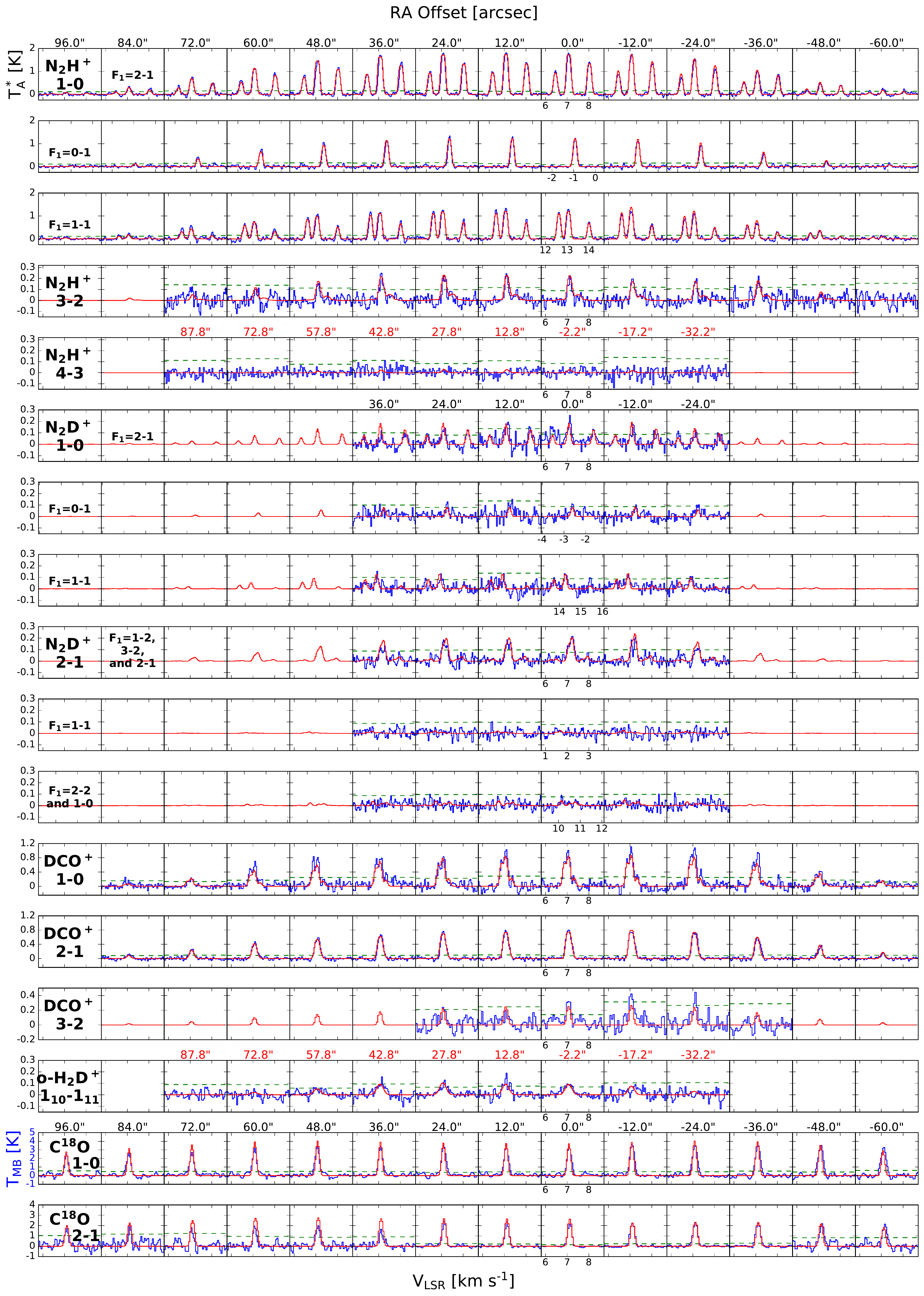}
	\caption{
    Spectral observations along the main horizontal cuts at $\Delta$Dec=$-$12\arcsec (IRAM/GBT) or $-$16\arcsec (JCMT) compared to our best-fit radiative transfer model.
    The blue spectra show the observational data, and the red spectra show the models. Each column corresponds to different horizontal offsets from the center of L\,1512 according to Fig.\,\ref{fig:pointing}. Each row shows a spectral line, except that the \NtHp (1--0), \NtDp (1--0) and (2--1) lines are split into three rows corresponding to their different $F_1$-transition groups. For \CetO (2--1), our data are supplemented with OTF spectra from previous works at large offsets.
    The green dashed lines indicate the three $\sigma$ noise level.}
    \label{fig:maincut_spectra}
\end{figure*}

\begin{figure*}
	\centering
	\includegraphics[width=18cm]{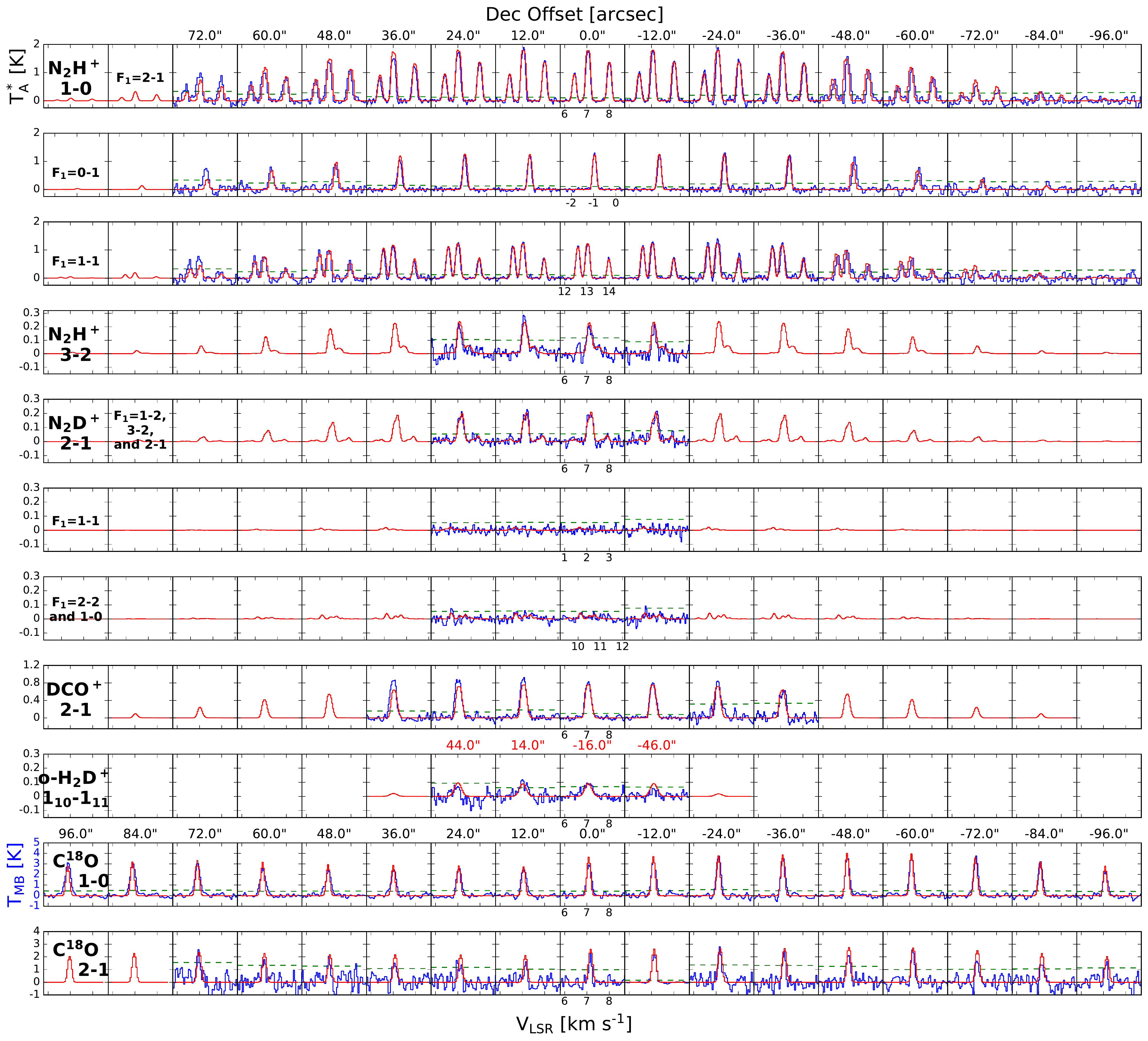}
	\caption{
    Spectral observations along the main vertical cuts at $\Delta$RA=0\arcsec (IRAM/GBT) or $-$2.2\arcsec (JCMT) compared to our best-fit radiative transfer model.
    For \NtHp (1--0) and \CetO (2--1), our data are supplemented with OTF spectra from previous works at large offsets.
    Other annotations are the same as in Fig.\,\ref{fig:maincut_spectra}.}
    \label{fig:vercut_spectra}
\end{figure*}

\begin{figure*}
	\centering
	\includegraphics[width=\textwidth]{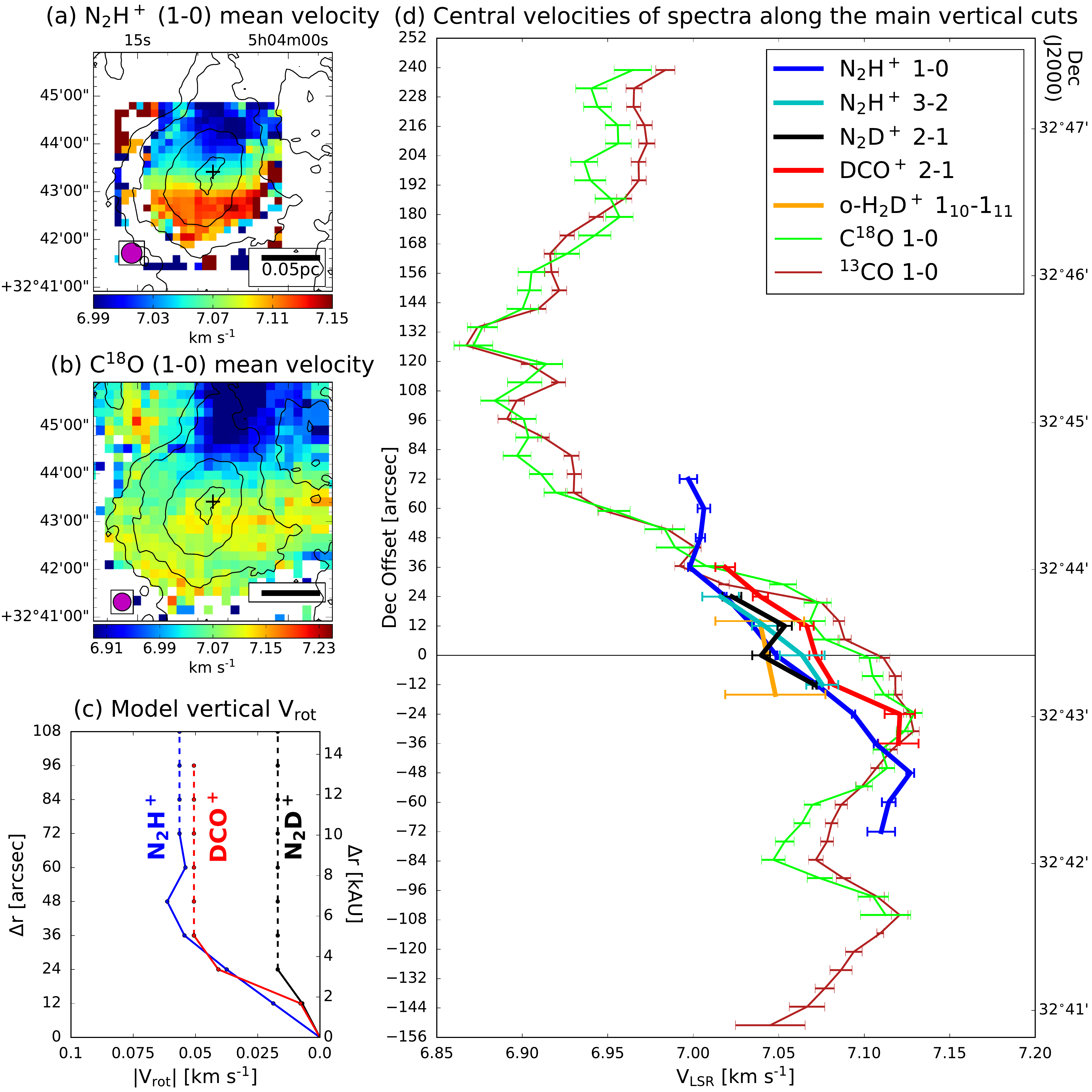}
	\caption{ 
    Velocity structure.
    Mean velocity maps of (a) the $F_1F$=2,3--2,1 component of \NtHp $J$=1--0 and (b) \CetO $J$=1--0 calculated within $V_{\rm LSR}$=[6.75 km s$^{-1}$, 7.50 km s$^{-1}$] and [6.0 km s$^{-1}$, 7.5 km s$^{-1}$], respectively. Contours from the the 850 \micron\,image are overlaid. 
    The rest of the symbols in the panels $a$ and $b$ have the same meaning as in Fig.\,\ref{fig:maps}.
    (c) Model vertical rotation velocities, $V_{\rm rot}$, against radius, $\Delta$r, adopted in the radiative transfer model.
    (d) Central velocities and their error bars of the spectra along the main vertical cuts. Two $y$-axes are vertical offsets to the center and declination, respectively.
    }
    \label{fig:velocity structure}
\end{figure*}

\subsection{Molecular emission lines}

Figure\,\ref{fig:pointing} shows the pointings of spectral line observations using IRAM 30 m, GBT, and JCMT HARP. Horizontally, our
IRAM/GBT observations at $\Delta{\rm Dec}=-12$\arcsec and our JCMT observation at $\Delta{\rm Dec}=-16$\arcsec have the widest
coverage across the L1512 core. Vertically, our IRAM/GBT observations at $\Delta{\rm RA}=0$\arcsec have the widest coverage, and
the JCMT observation closest to it is at $\Delta{\rm RA}=-2$\farcs2. We used these cuts (hereafter referring to them as the main
horizontal cuts and the main vertical cuts) to analyze the physical and chemical properties of L1512 (Sect.\,\ref{sec:analysis}).
Figures\,\ref{fig:maincut_spectra} and \ref{fig:vercut_spectra} show the spectra of each emission line in the $T_{\rm A}^*$ scale (except for \CetO (1--0) in the $T_{\rm MB}$ scale)
along the main horizontal  and  vertical cuts, respectively.

Since L1512 is a quiescent starless core, each hyperfine component of \NtHp (1--0) and \NtDp (1--0) is well separated with line widths of $\sim$0.2 km s$^{-1}$.
Along the main horizontal cuts, 
the coverage of the \NtHp (1--0) spectra includes the whole starless core 
with the intensity peak toward the central region.
Likewise, the spectra of \NtHp (3--2), \NtDp, and \DCOp also peak toward the central region, which suggests that these four cations trace the dense core region.
In addition, our highest observed transition of \NtHp, the $J$=4--3 line, shows no signal according to the 3$\sigma$ criterion with a rms noise level of $\sim$30 mK because of the 
low kinetic temperature in the core region and low density compared to the requested critical density for this transition (two orders of magnitude lower).
Among the four cations, the ortho-\HtDp (\oHtDpgrd) line is detected in a smaller spatial region ($\Delta{\rm RA}=-$17.2\arcsec to 42.8\arcsec),
which indicates that ortho-\HtDp 
traces the innermost region of the L\,1512 core.
On the other hand, the intensity of the \CetO\ (1--0) and (2--1) spectra are relatively constant across the inner core,
which is a clear indication that \CetO is depleted across the whole core.

Along the main vertical cuts, Fig.\,\ref{fig:vercut_spectra} shows that \NtHp, \NtDp, and \DCOp are also depleted in the innermost
region. Their spectral peaks drop between the region from $\Delta{\rm Dec}=-$24 to $+$12\arcsec for which the intensity decrements range from 0.04 to 0.15 K with $\sim$2--5$\sigma$ significant levels, while they should increase to follow the increase of column density toward the center if
there were no depletion. Similarly, the ortho-\HtDp spectra show a roughly constant intensity across the peak, which is a sure sign of
depletion since the line is optically thin and should be roughly proportional to the H$_2$ column density. The depletion effects
are also found in the integrated intensity maps of \NtHp (1--0) and \CetO (1--0) shown in Figs.\,\ref{fig:maps}h and i. The \CetO 
emission clearly shows depletion toward the center of the cloud because the peak emission is situated on the outskirts, while an
offset between \NtHp emission peak and dust peak suggests that \NtHp also depletes in the innermost region. 

\subsection{Velocity structure}

Figures\,\ref{fig:velocity structure}a and b show the mean velocity maps of L\,1512, which reveals a velocity gradient in the
north-south direction. The constant gradient inside the core itself is compatible with a rigid-like rotation feature or could be due to an oscillation of the whole filament. Figure\,\ref{fig:velocity structure}d shows the central
velocity of each spectral line along the main vertical cuts fitted with the CLASS Gaussian  fitting task. The hyperfine
structures of the \NtHp, \NtDp, and \DCOp rotational lines are taken into account for the fitting. We see that the \NtHp (1--0)
emission appears to have a rather uniform velocity gradient of 2.26\,$\pm$\,0.04 km\,s$^{-1}$\,pc$^{-1}$ between $\Delta{\rm
Dec}=-48$\arcsec and $+$36\arcsec, which is compatible with a rigid body rotation. The \NtHp (3--2), \NtDp (2--1), \DCOp (2--1), (and
possibly ortho-\HtDp (\oHtDpgrd) although its velocity measurement is hampered by a low signal-to-noise ratio) emissions also have similar
dynamical behaviors. 
The \CetO (1--0) and \thtCO (1--0) data from \citet{Falgarone98} partially follow the ionic species velocity drift but
with an offset of $\sim$50 m\,s$^{-1}$ toward the center. Because the \thtCO emission traces a more extended region than
\CetO, the similarity between the \CetO and \thtCO emissions might be due to \CetO depletion in the core region and its emission
being dominated by the L\,1512 envelope.  The CO isotopolog redshift by $\sim$50 m\,s$^{-1}$ toward the center could be explained by a contraction of the 
envelope, if the line is optically thick enough to mask the corresponding blueshifted component that we would expect on the back side of the cloud; however, this seems difficult to justify for the \CetO (1--0) transition.  The other possible explanation is that the envelope is experiencing an oscillation in which the core is coupled to it with a delay and is presently toppling over. 
This kind of oscillation has been proposed to explain the Orion  and the G350.54+0.69 filaments \citep{Stutz18,LiuHL19} and has also been invoked for L\,1512 itself \citep{Lee01, Sohn07, Lee11}.

\section{Analysis}\label{sec:analysis}

Our first step is to analyze the dust absorption maps from 1.2 \micron to 4.5 \micron to estimate the total column density of
gas+dust all over the cloud. We converted this map into a sphere on first-order approximation to analyze the line emissions. We
adopted a correspondingly 1D spherically symmetric non-LTE radiative transfer code \citep{Bernes79, Pagani07} to reproduce
our observed emission line spectra. Since the shape of L\,1512 is globular, we can assume that its distributions of volumetric
number density ($n_{\rm H_2}$), gas kinetic temperature ($T_{\rm kin}$), volumetric relative abundances with respect to \Ht of the observed
species ($X_{\rm species}=n_{\rm species}/n_{\rm H_2}$) and turbulent velocity ($V_{\rm turb}$) can be approximated by an
onion-shell structure composed of multiple concentric homogeneous layers. We note however that the cloud is slightly
dissymmetric and we used two slightly different models to describe the east and west sides of the cloud. In addition, rotational velocity field ($V_{\rm rot}$)
and radial velocity field ($V_{\rm rad}$) can also be applied to each layer. We chose the widths of the layers to be 12\arcsec
(=1,680 AU), which is the smaller spacing of the two pointing-grids (Fig.\,\ref{fig:pointing}). The advantage is that we can
determine the physical parameters and chemical abundances at each layer sequentially from the outermost to the innermost layer by
sampling sightlines of progressively decreasing radius along a cut across L\,1512.

Our procedure to determine $n_{\rm H_2}$, $T_{\rm kin}$, $X_{\rm species}$, and $V_{\rm turb}$ of each layer is as follows. First,
the $n_{\rm H_2}$ profile was determined independently from an extinction map. Second, we used \NtHp data to determine their $T_{\rm
kin}$, $X$(\NtHp), and $V_{\rm turb}$ profiles. Third, we assumed that all the other species have the same $T_{\rm kin}$ and
$V_{\rm turb}$ profiles as those of \NtHp to determine the abundance profiles of \NtDp, \DCOp, and ortho-\HtDp. Given the
abundance profiles of the various ions, we adopted a pseudo time-dependent chemical model \citep{Pagani09b} to estimate the lifetime scale of each layer
with our derived  $n_{\rm H_2}$ and $T_{\rm kin}$, and we also derived the abundance profiles of their parent species, CO and N$_2$.

\subsection{Visual extinction}

We adopted the Near-Infrared Color Excess Revisited technique \citep[NICER;][]{Lombardi01} to derive the visual extinction, \Av,
with color excesses of stars from NIR and MIR bands. Since the NIR data preferentially trace the diffuse region, we used the
$R_{\rm V}=3.1$ model from \citet{Weingartner01} to derive the  \Av map  of the envelope (Fig.\,\ref{fig:extinction}a) using \J, \H, and \Ks
bands. On the other hand, the MIR data preferentially trace the dense region. We alternatively used the $R_{\rm V}=5.5$ case B model from
\citet{Weingartner01} to derive \Av map of the core region (Fig.\,\ref{fig:extinction}b) using \H, \Ks, and IRAC2 bands. The case
B model is an improved variant of the $R_{\rm V}=5.5$ model from \citet{Weingartner01}, but characterizes 
the denser region better (by including spherical grains up to 10 \micron in size), although not perfectly \citep{Cambresy11,Ascenso13}. Thanks to the deep sensitivity
of the CFHT and Spitzer observations, the density of stars across the cloud core is high enough to convolve the reddening data with a 2D Gaussian profile
of half-power beam width (HPBW) as small as 30\arcsec to construct the maps. The dense core map traces the central region better than the
diffuse envelope map but could not properly trace the outer low-\Av region, and vice versa. In order to combine these two maps, we
calculated the azimuthally-averaged \Av profiles of the  envelope and core  maps shown in Fig.\,\ref{fig:extinction}d
from the 1\arcmin horizontal strip across the center (Figs.\,\ref{fig:extinction}a,b) to avoid the cometary tail. We could see
that the two profiles merge at $\sim$2.5 mag and become similar in the outer region. Therefore, an \Av threshold of 2.5 mag is
used to combine the central region from the core map and the outer region from the envelope map, and the result is shown in
Fig.\,\ref{fig:extinction}c.

\subsection{Density profile}
The volume density ($n_{\rm H_2}$) profile of L\,1512 is modeled with the spherically symmetric Plummer-like profile,
\begin{equation}
    n_{\rm H_2}(r) =
    \begin{dcases}
        \frac{n_0}{1+\big(\frac{r}{R_0}\big)^\eta} & \text{if $r \leq R_{\rm edge}$}\\
        0 & \text{otherwise,}
    \end{dcases}
\end{equation}

where $n_0$ is the central density, $R_0$ is the characteristic radius, $\eta$ is the power-law index of $n_{\rm H_2}$ profile as
$r\gg R_0$, and $R_{\rm edge}$ is set as 320\arcsec ($=$44,800\,AU) to cover the whole core. 
Since the contribution from the diffuse envelope is small compared with the dense core, we simply adopted the $R_{\rm V}=5.5$B model for the conversion of the extinction, \Av, into column density, $N_{\rm H_2}$.
Therefore we obtain a relation between \Av and $N_{\rm H_2}$ \citep{Bohlin78},

\begin{equation}
N_{\rm H_2}/A_{\rm V}=5.3\times10^{20} \text{ cm$^{-2}$\, mag$^{-1}$.}
\label{equ:Av}
\end{equation}

\begin{figure*}
	\centering
    \includegraphics[scale=0.42]{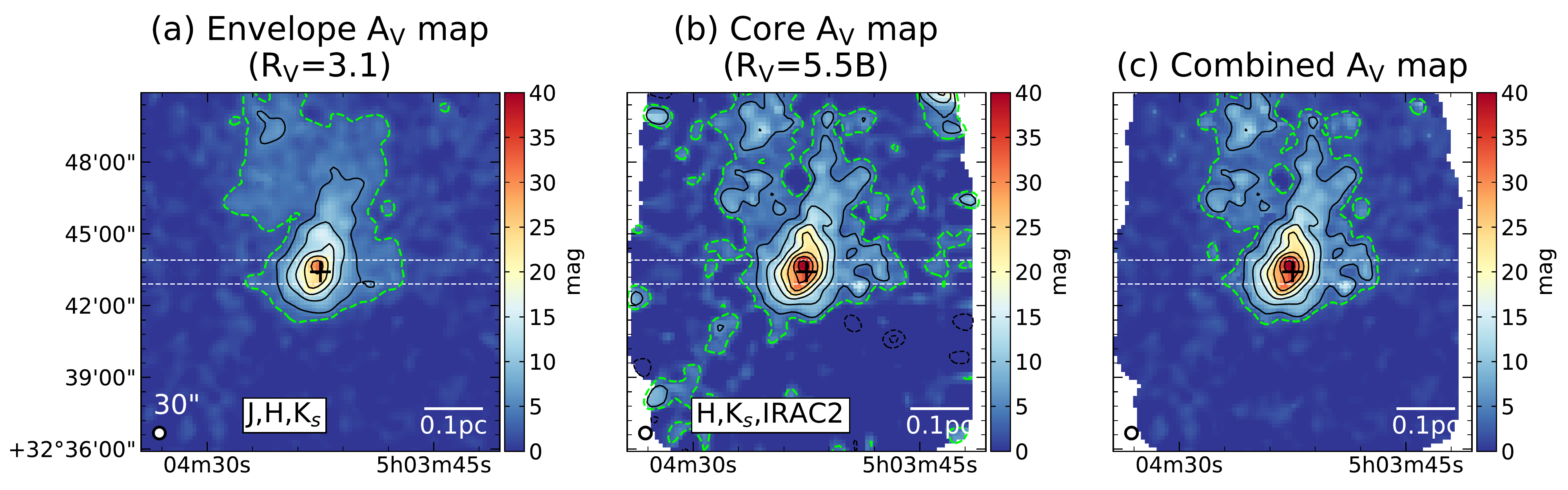}
    \includegraphics[scale=0.4]{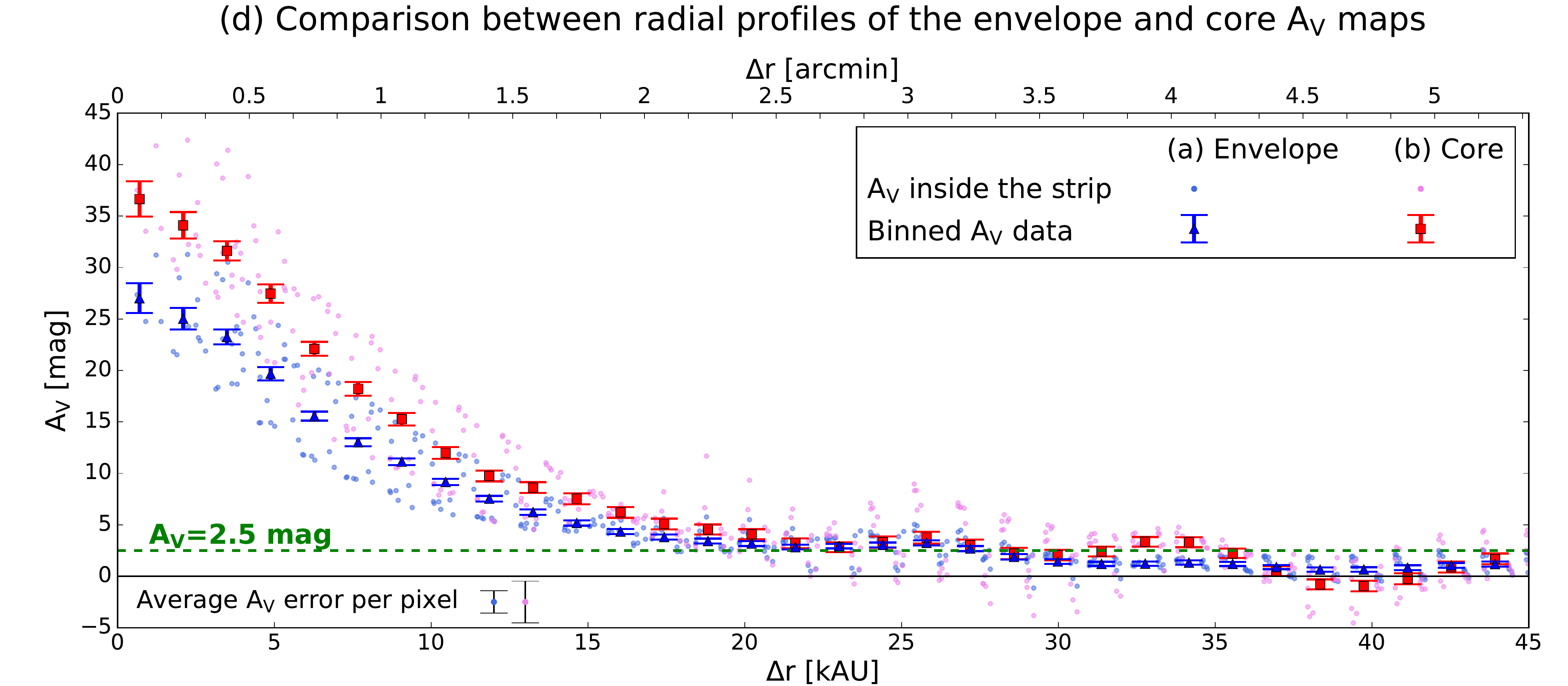}
    \caption{
    Visual extinction maps and profiles.
    (a) Envelope $A_{\rm V}$ map derived from \J, \H, and \Ks bands with the $R_{\rm V}=3.1$ model. 
    (b) Core $A_{\rm V}$ map derived from \H, \Ks, and IRAC2 bands with the $R_{\rm V}=5.5$B  model.
    (c) Combined $A_{\rm V}$ map merged from the previous maps with boundaries at $A_{\rm V}=2.5$ mag. Each map is convolved with a Gaussian beam size of 30\arcsec.
    (d) Comparison between $A_{\rm V}$ profiles inside the strips of panel $a$ and $b$.
    The $A_{\rm V}$ values at each pixel are denoted in light blue and pink dots, while their average errors are indicated in the bottom left corner.
    The $A_{\rm V}$ profiles are averaged in azimuth with radial 10\arcsec\ bin and shown in blue triangles and red squares with their error bars.
    The green dashed lines in the four planes indicate the $A_{\rm V}$ threshold of 2.5 mag.}
    \label{fig:extinction}
\end{figure*}

In order to derive $n_0$, $R_0$ and $\eta$ in the volume density model of L\,1512, we
produced an \Av model map by (1) integrating a Plummer-like $n_{\rm H_2}$ model along the line of sight and convolving it with the
same beam size as the combined \Av map (Fig.\,\ref{fig:extinction}c) to obtain an $N_{\rm H_2}$ model map, and then (2) derive an
\Av model map using Eq.\,\ref{equ:Av}. Therefore, we could perform a fitting on the azimuthally-averaged profiles of the \Av model
map and the combined \Av map with the Levenberg-Marquardt algorithm included in the Python \texttt{scipy.optimize.curve\_fit}
function.

\begin{figure*}
	\centering
    \includegraphics[scale=0.42]{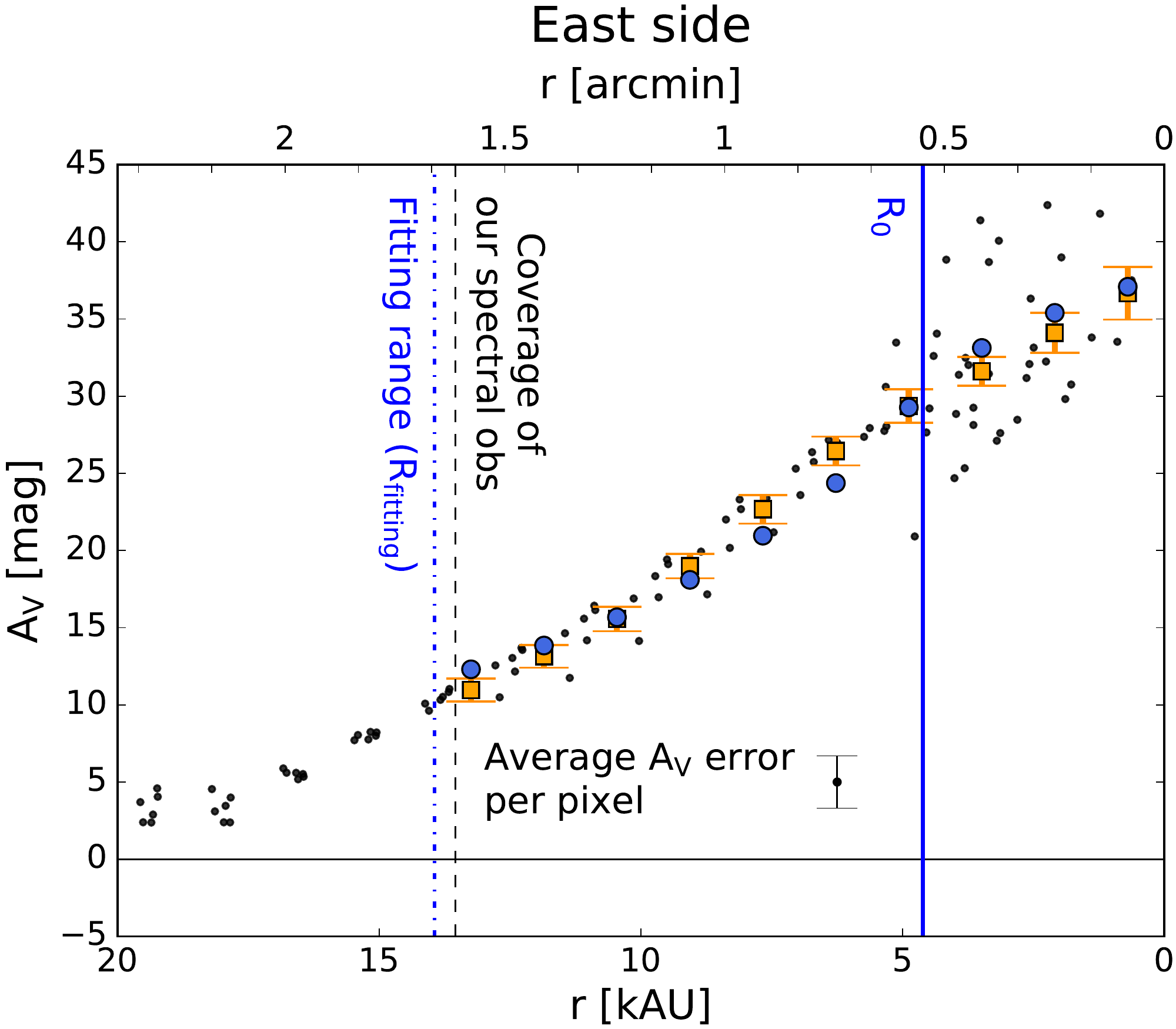}\includegraphics[scale=0.42]{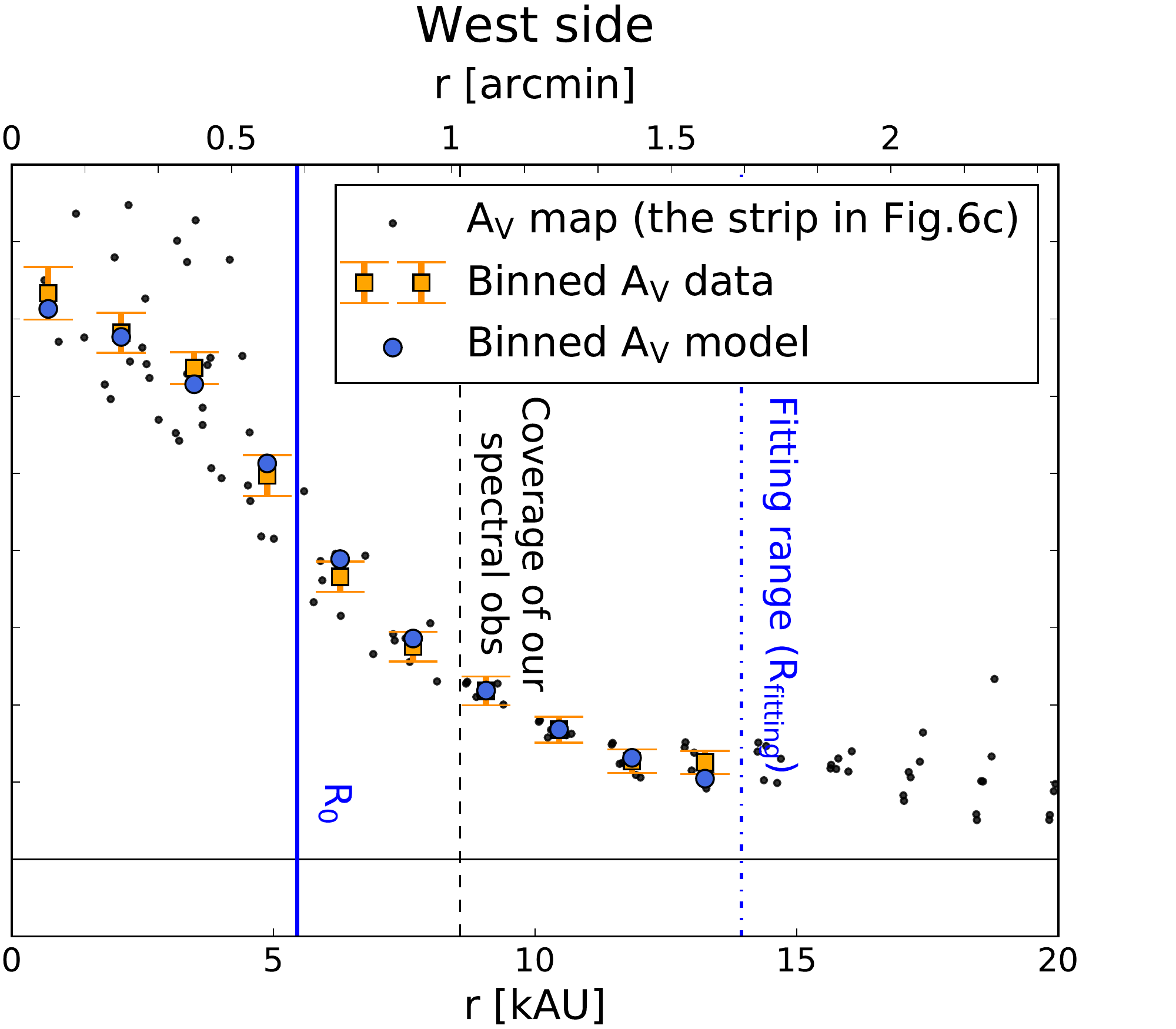}
    \caption{
    Plummer-like density profile fittings.
    The black dots are the \Av values at each pixel in the horizontal strip shown in Fig.\,\ref{fig:extinction}c but these are separated between the east and west sides.
    The orange squares with error bars show the averaged \Av profiles, which are azimuthally binned with radial 10\arcsec\ bin.
    The blue circles show the best-fitted \Av models for the east and west sides.
    The \Av models are derived from spherically symmetrical Plummer-like density profiles.
    The best-fitted characteristic radii, $R_{0}$, and the fitting ranges, $R_{\rm fitting}$ of the \Av models are denoted in blue solid and dash-dotted lines, respectively.
    The coverage of our spectral observations are denoted in black dashed lines.}
    \label{fig:Av_fitting}
\end{figure*}

\begin{figure*}[!ht]
	\centering
	\includegraphics[scale=0.5]{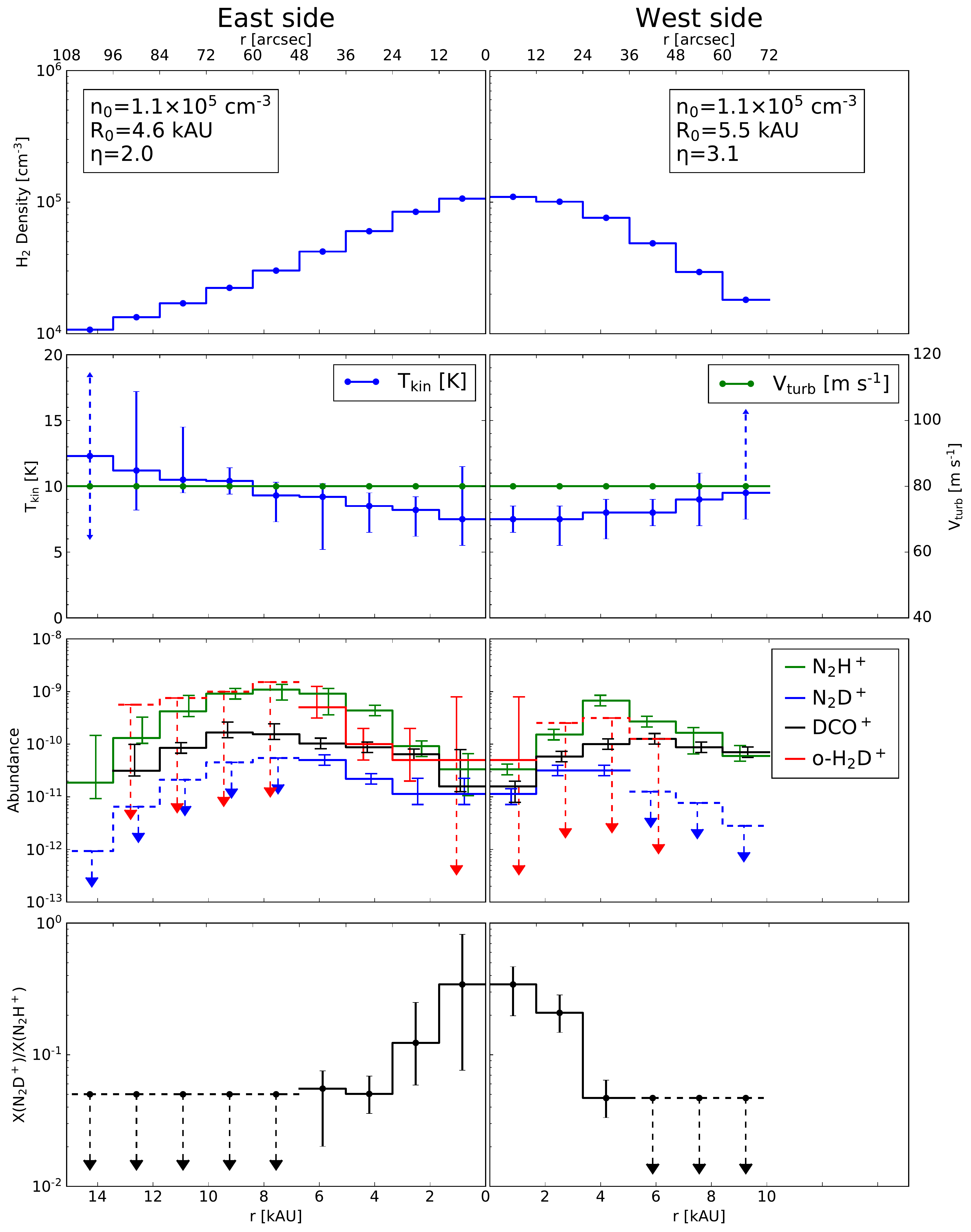}
	\caption{
    Physical and abundance profiles.
    The Plummer-like density ($n_{\rm H_2}$) profiles are derived from the combined extinction map (Fig.\,\ref{fig:extinction}c), where the best-fit parameters are annotated.
    The profiles of kinetic temperature ($T_{\rm kin}$), turbulent velocity ($V_{\rm turb}$), abundances ($X_{\rm species}$), and the abundance ratio of \NtDp/\NtHp are the best-fit results from the non-LTE radiative transfer calculations.
    }
    \label{fig:rad_model}
\end{figure*}

Figure\,\ref{fig:Av_fitting} shows the fitting results for the east and west sides of L\,1512 using the \Av computed in the strip shown in Fig.\,\ref{fig:extinction}c, which spatially covers the main horizontal cuts.
In order to ensure that the volume density models of both sides agree with each other at the center,
$n_0$ was previously fitted by an averaged profile from both sides. 
Then $n_0$ was fixed in the individual fitting for each side.
These determined parameters and volume density profiles are shown in the first row in Fig.\,\ref{fig:rad_model}.
We  see that the horizontal $n_{\rm H_2}$ profile is skewed in such a way that it is steeper on the west side  than the other side. 
The volume densities drop to $\sim$10$^4$ cm$^{-3}$
at both edges of our spectral observation coverage. 

\subsection{Radiative transfer applied to the onion-shell model}\label{sec:analysis_rad}

We assumed an onion-shell model comprised of nine and six 12\arcsec layers in thickness on the east and west sides, respectively. Then we
reproduced the spectra of \NtHp (1--0, 3--2, and 4--3) along the main horizontal cuts with the 1D spherically symmetric non-LTE
radiative transfer code \citep{Pagani07} by varying $T_{\rm kin}$, $X$(\NtHp), $V_{\rm turb}$, $V_{\rm rot}$, and $V_{\rm rad}$ at
each layer. Here we used the para-H$_2$--\NtHp collisional coefficients instead of the original He--\NtHp hyperfine collisional
coefficients \citep{Lique14}. We also used specific para-H$_2$--\NtDp collisional rate coefficients for the deuterated isotopolog (see Appendix \ref{N2D+collcoef}).  
One of the important features of this code is that the line overlap between hyperfine transitions at close
frequencies is taken into account to calculate the correct spectral line shape and excitation. Among the input parameters, $V_{\rm turb}$, $V_{\rm
rot}$, and $V_{\rm rad}$ are determined from observations. We set both $V_{\rm rot}$ and $V_{\rm rad}$ to zero, since the
L\,1512 core has no obvious infall/expansion motion and only a vertical rotational component, while our model follows the RA (horizontal) structure. 
For $V_{\rm turb}$, we found that a uniform turbulent velocity of 80 m\,s$^{-1}$ could reproduce the spectral line widths along the cuts. Afterward, we performed an
iterative spectral fitting procedure to obtain $T_{\rm kin}$ and $X$(\NtHp)  profiles, taking care to eliminate nonphysical solutions such as parameter oscillations 
between layers, even if they provide better $\chi^2$.

An important feature of our approach is that instead of converting the data into the $T_\mathrm{MB}$ scale for direct
comparison with the non-LTE model, which is absolutely not valid for extended regions such as starless clouds, we reproduce the
observations by computing the telescope coupling to the sky, based on its main beam, error beams, and forward scattering and spillover
efficiency ($\eta_\mathrm{fss}$)\footnote{http://www.iram.es/IRAMES/mainWiki/Iram30mEfficiencies}.  We then obtained our model predictions directly
calibrated in the $T_\mathrm{A}^*$ scale, whereas the $T_\mathrm{MB}$ scale is overcorrected for low $\eta_\mathrm{MB}$ values such as those
found for the IRAM 30 m in the 1.3--0.8 mm range since the error beams pick up a substantial fraction of the signal when the
source is extended. This leads to wrong line ratios and overestimates of the temperature and/or density of the cloud and also has
an impact on the species abundance. 

Once the density, temperature, and kinematic properties of the onion-shell model were constrained by the \NtHp modeling, we
reproduced the spectra of \NtDp (1--0, and 2--1), \DCOp (1--0, 2--1, and 3--2), and ortho-\HtDp (\oHtDpgrd) by varying only their
abundances and assuming that they share the same physical parameters as \NtHp. Similarly, $X$(\NtDp), $X$(\DCOp), and $X$(ortho-\HtDp)
profiles were obtained using the above iteratively spectral fitting procedure. Since we lacked \NtDp observations at large
radii, we set upper limits of $X$(\NtDp) in the outer layers from $X$(\NtHp) multiplied by the minimum observed
$X$(\NtDp)/$X$(\NtHp$)=0.05$. For the non-detection observations of ortho-\HtDp, we also estimated their abundance upper limits.

Figure\,\ref{fig:rad_model} shows the profiles of $T_{\rm kin}$, $V_{\rm turb}$, abundances of the above four cations, and 
the \NtHp fractionation, $X$(\NtDp)/$X$(\NtHp). 
The above best-fit profiles are also numerically listed in Table\,\ref{tab:rad_model}.
The error bar in each layer is determined by the range of quantity, where
\begin{equation}\label{equ:error}
\Delta \chi^2=\chi^2(\text{quantity})-\chi^2(\text{the best-fit quantity})\leq1.    
\end{equation}
We only determined the upper limit of $X$(ortho-\HtDp) toward the innermost layer.
A higher signal-to-noise ratio of ortho-\HtDp would be needed to derive its lower limit
because the volume of the innermost layer is the smallest and its contribution to the emergent intensity is also small.

Finally, the best-fit modeled spectra of the four cations along the main horizontal cuts are shown in red in
Fig.\,\ref{fig:maincut_spectra}. The fit with the observed spectra is remarkably good. To see if the observed spectra along
the main vertical cut could also be reproduced by our onion-shell model, we simply assumed that the physical parameters and
abundance profiles were the same as those of the east side except for $V_{\rm rot}$ profile that is now not zero. Since the
L\,1512 core has a vertical motion compatible with a solid-body rotation, we adopted 
the $V_{\rm rot}$ profile of \NtHp, \NtDp, and \DCOp shown in Fig.\,\ref{fig:velocity structure}c, which
averages the central velocities of the vertical spectra (Fig.\,\ref{fig:velocity structure}d) between the north and south sides. 
For ortho-\HtDp, we simply adopted the $V_{\rm rot}$ profile of \NtHp because of its weak detection.
The above $V_{\rm rot}$ profiles are given at layer interfaces and the rotational velocity inside a layer is linearly interpolated
between the two given velocities at the inner and outer sides of that layer. The modeled spectra of the four cations are shown in red in Fig.\,\ref{fig:vercut_spectra}, and are again in good agreement with the observations. Our results
suggest that the 1D spherically symmetric assumption is globally valid for L\,1512.

\subsection{Time-dependent chemical model}\label{sec:analysis_chem}

We adopted a pseudo time-dependent gas-phase chemical model \citep{Pagani09b} to reproduce the deuteration ratio, $X$(\NtDp)/$X$(\NtHp), 
and abundances of \NtHp, \NtDp, \DCOp, and ortho-\HtDp in each layer.
In the pseudo time-dependent approach, the physical properties ($n_{\rm H_2}$, $T_{\rm kin}$) are kept constant while the chemical abundances ($X_{\rm species}$) evolve. 
We considered a chemical network, where (1) different spin states (ortho, para, and meta) of \Ht and \Hthp, and their isotopologs are all included, and (2) heavy molecules are totally depleted except for CO and N$_2$ which are partially depleted.

The initial ortho-\Ht/para-\Ht ratio is assumed to be their statistical weights of 3:1
and \Hthp is formed later via the cosmic ray ionization of \Ht. After the ortho/para ratio (OPR) of H$_2$ has dropped to values low enough to prevent destruction of \HtDp by ortho-\Ht (OPR
$\lesssim$ 1\% or less; \citealt{Pagani92b,Pagani09b}), the deuterium fractionation of \Hthp  enhances in the low $T_{\rm kin}$ and 
highly depleted environment, forming \HtDp, and subsequently \DtHp and D$_3^+$. Then \Hthp and its isotopologs transfer
protons/deuterons to CO and N$_2$ and produce HCO$^+$, \DCOp, \NtHp, and \NtDp. HCO$^+$ is not a suitable tracer for
starless cores because its  low rotational transition emission is optically thick and this ion is present in both the envelope
and the core (which is also true for H$^{13}$CO$^+$ and HC$^{18}$O$^+$), while ortho-\HtDp, \DCOp, \NtHp, and \NtDp emissions are
optically thinner and most importantly are confined to the core \citep{Pagani12}. 
Since we evaluated the quasi-instantaneous conversion of CO and N$_2$ into the aforementioned ions, we can assume that their
abundances are in equilibrium with these ions locally. Therefore, the gas-phase  abundances of these two neutral molecules are 
free parameters in our chemical model. The other free parameters are the average grain size ($a_{\rm gr}$) and the cosmic ray
ionization rate ($\zeta$).

\begin{figure*}
	\centering
	\includegraphics[scale=0.35]{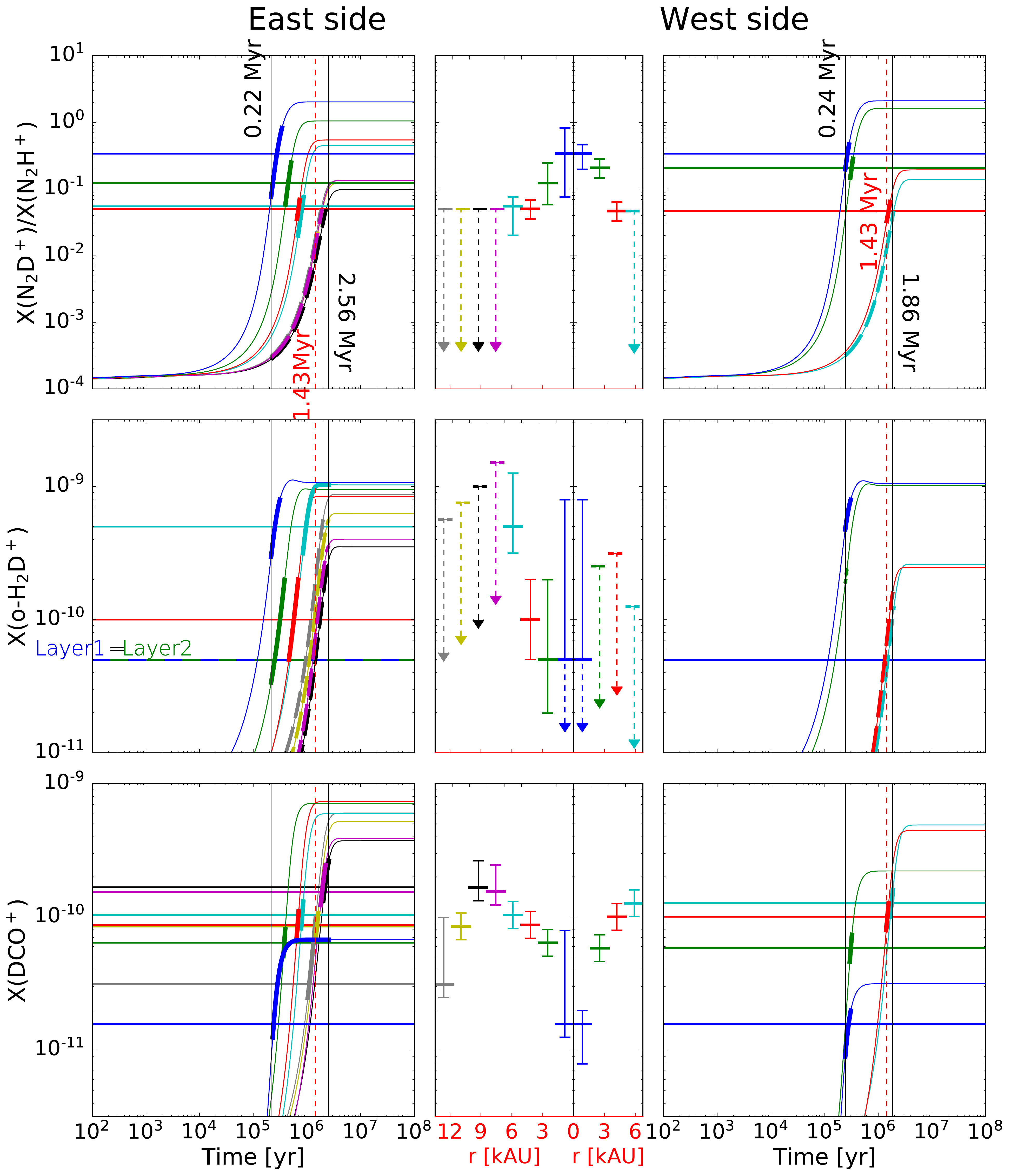}
	\caption{
    Chemical modeling of the abundance ratio of \NtDp/\NtHp and the abundances of
    ortho-\HtDp and \DCOp for each layer.
    The left and right columns show the chemical solutions, while the middle column shows the profiles.
    Each color represents a layer.
    In the chemical solution plots, the horizontal solid lines are the observationally derived  $X$(\NtDp)/$X$(\NtHp), $X$(ortho-\HtDp), and $X$(\DCOp).
    The growing curves are the chemical solutions calculated with an average grain size ($a_{\rm gr}$) of 0.1\,\micron and a cosmic ray ionization rate ($\zeta$) of 10$^{-17}$\,s$^{-1}$.
    The two black vertical lines in each panel indicate a time range for which the model values cross the observations within their error bars and where the model curves become thicker. 
    Thick dashed lines denote the model range that is both lower than the observed upper limit for that layer and within the global time range of the model. 
    The red dashed vertical lines indicate the lower limit on the lifetime scale of L1512.}
    \label{fig:chem_model}
\end{figure*}

Figure\,\ref{fig:chem_model} shows the chemical model solutions of $X$(ortho-\HtDp), $X$(\NtDp)/$X$(\NtHp), and $X$(\DCOp) for each layer
toward the east and west sides. 
Typical $\zeta$ of 10$^{-17}$ s$^{-1}$ was adopted throughout the chemical model. 
For $a_{\rm gr}$, although \citet{Steinacker15} find that the maximum spherical grain size can reach $\sim$0.5\,\micron in this source, the majority of the grains are still small and the larger grains have fluffy surfaces \citep[e.g.,][]{Tazaki16}, therefore limiting the decrease of the total grain surface cross section. 
The impact on the chemistry and the freeze-out time should be negligible compared to 
the standard 0.1\,\micron grain size usually adopted in chemical models. In each layer, $X$(CO) and $X$(N$_2$) were adjusted to make the chemical solutions of the four cations fit the observed abundances
(or remain below the observed upper limits) simultaneously. We found that the time ranges for all layers meeting the observations
are 0.2--2.6 Myr for the east side and 0.2--1.9 Myr for the west side.

\begin{figure*}
	\sidecaption
	\includegraphics[width=12cm]{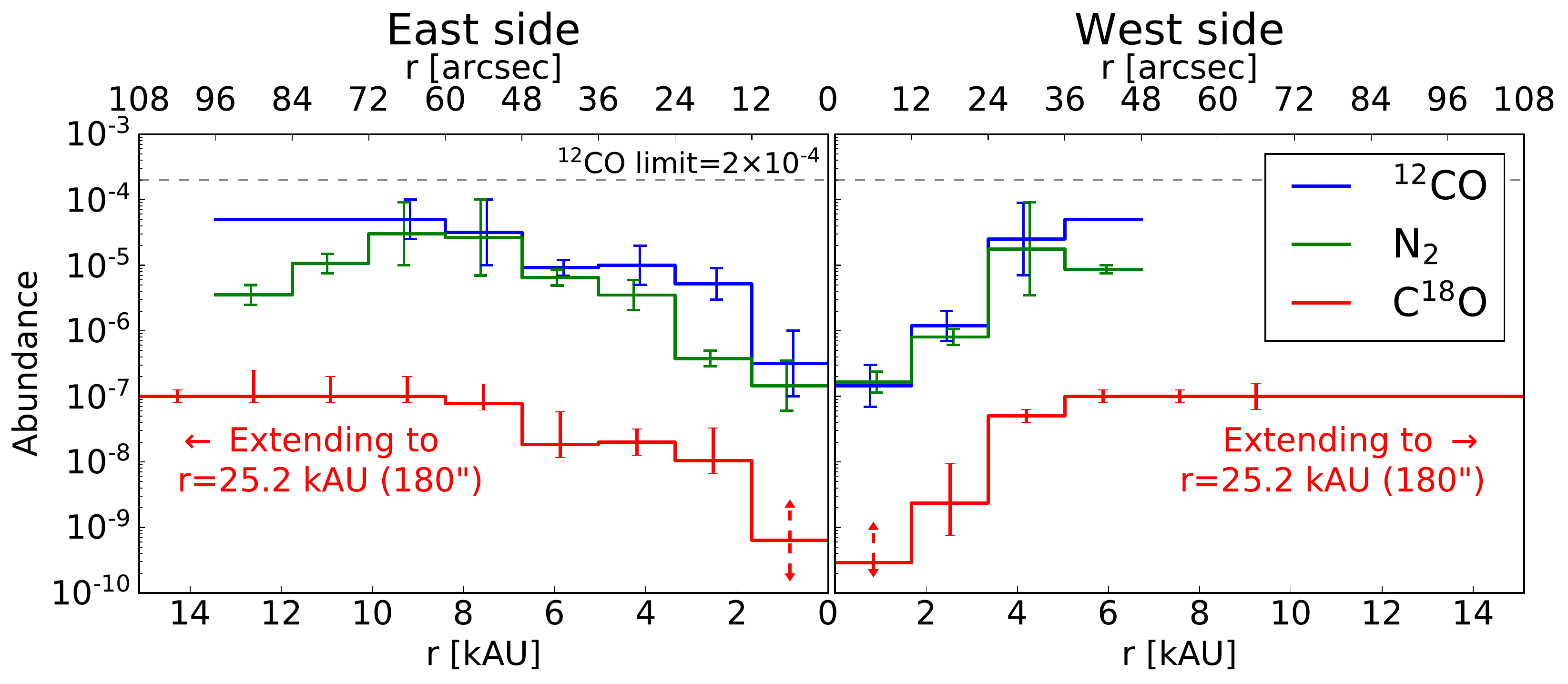}
	\caption{
    \twvCO, N$_2$, and \CetO abundance profiles.
    The \twvCO and N$_2$ abundance profiles are the best-fit results from the chemical modeling. The \CetO abundance profile is obtained by assuming a \twvCO/\CetO 
    abundance ratio of 500 or a constant value of 1.0 $\times 10^{-7}$ extending to 25.3\,kAU.}
    \label{fig:chem_model_CO_N2}
\end{figure*}

Figure\,\ref{fig:chem_model_CO_N2} shows the derived CO and N$_2$ abundance profiles, while their values are listed in Table\,\ref{tab:rad_model}.
Although the isotopologs of CO are not considered
in our chemical network, we can obtain the $X$(\CetO) profile through a $^{12}$C$^{16}$O/$^{12}$\CetO abundance ratio in the layers in which \CetO cannot be measured directly.
In the outer region, where radiative transfer modeling is applicable, we find that the \CetO abundance is $1\times10^{-7}$ up to a radius of 3\arcmin (=25,200 AU).
Since $X$(\CetO) is limited to $1\times 10^{-7}$ toward the sixth eastern layer,
where the \twvCO abundance reaches its maximum of $5\times 10^{-5}$,
we find an isotopolog ratio of $^{12}$C$^{16}$O/$^{12}$\CetO = 500. This is compatible with the local $^{16}$O/$^{18}$O ISM ratio of 557$\pm$30 \citep{Wilson94}.
As there is no reason to vary this ratio across the cloud, we obtained the inner \CetO abundance profile between the fifth eastern and the third western layers by 
applying the same ratio of 500 to $X$(\twvCO) profile.
On the other hand, the outer \twvCO abundance profile is kept as high as $5\times 10^{-5}$ 
because $X$(\twvCO) is poorly determined from the chemical model when 
both $X$(\NtDp)/$X$(\NtHp) and $X$(ortho-\HtDp) are only upper limits.
As a result, we were able to reproduce the \CetO spectra via the non-LTE radiative transfer calculation. 
Owing to the nonsymmetrical velocity shift of \CetO, we adjusted the velocity offset of the modeled spectra individually on each spectrum based on the observations. The impact on the line shape is too small to be detectable.
The northern \CetO abundance profile (positive offsets) is different from the east-west profile and had to be decreased by a factor of 2--16 to fit the observations. This suggests a larger CO depletion toward the northern tail.
It turns out that the observations can be fitted by our model results of \CetO (1--0) and (2--1) along the main horizontal cut (Fig.\,\ref{fig:maincut_spectra}) and
main vertical cut (southward only, Fig.\,\ref{fig:vercut_spectra}) computed with the abundances derived from the chemical model,
thereby validating our chemical model results.

\section{Discussion}\label{sec:discussion}

As shown in Sect.\,\ref{sec:analysis}, we built an onion-shell model of L\,1512 to represent its physical structure ($n_{\rm
H_2}$, $T_{\rm kin}$, and $V_{\rm turb}$ are shown in Fig.\,\ref{fig:rad_model}; $V_{\rm rot}$ in Fig.\,\ref{fig:velocity
structure}c) and chemical abundances (\NtHp, \NtDp, \DCOp, and ortho-\HtDp are shown in Fig.\,\ref{fig:rad_model}; CO, \CetO, and
N$_2$ in Fig.\,\ref{fig:chem_model_CO_N2}). Our model reproduces all the observed spectra along the main horizontal and vertical
cuts (Figs.\,\ref{fig:maincut_spectra} and \ref{fig:vercut_spectra}) via the non-LTE radiative transfer calculations. In this
section, we first compare our findings with other studies and then address the lifetime of L\,1512.

\subsection{Density and kinetic temperature}

By integrating the Plummer profile to the edge of the cloud where it merges in extinction with the background
($\sim$320\arcsec from center), we find a peak $N_{\rm H_2}$ of $2.3\times10^{22}$ cm$^{-2}$ toward the center of L\,1512, which is 
higher than previously reported ($0.8-1.3\times10^{22}$ cm$^{-2}$; \citealt{Lee03, Lippok13, Lippok16}). From
the non-LTE modeling of both \CetO and \NtHp we also find 2.1 $\times10^{22}$ cm$^{-2}$ in a 180\arcsec radius, which is a lower
limit since the cloud is truncated to the \CetO emission extension and is consistent with the Plummer profile. This discrepancy is due to our different approach to derive the total column density. We use dust extinction, which is independent
of the dust temperature ($T_{\rm dust}$), and \NtHp modeling for which the uncertainty on the collisional coefficients is extremely
small compared to the uncertainty on dust properties in emission at long wavelengths. On top of dust properties uncertainties, it
is well known that the problem of quantifying dust in emission is degenerate. This is because of the unconstrained temperature variation
along the line of sight that can hardly be disentangled from dust property variations and the nonlinearity of the blackbody
intensity with temperature for the combination of wavelengths and temperatures of concern in starless cores \citep{Pagani15}. Indeed,
\citet{Lee03} and \citet{Lippok13, Lippok16} performed their estimations with dust thermal emission by assuming a constant
$T_{\rm dust}$ of 10 K or using a decreasing $T_{\rm dust}$ toward the core center. However, their $N_{\rm H_2}$ are less than
$\sim$50\% of our result. \citet{Pagani15} suggest that if only dust continuum data are used, $T_{\rm dust}$ can be
overestimated and consequently  the column density of starless cores can be underestimated. This is because warmer dust emission can
easily dominate over cold dust emission in the spectral energy distribution fitting. 

For the kinetic temperature, we find a central $T_{\rm kin}$ of 7.5$\pm{1}$\,K on the west side, which is better constrained
that the central temperature for the east side (7.5$_{-2}^{+4}$ K) . We also note that the second central layer on the east side
is better constrained too ($8.2_{-2}^{+1}$ K). Thus, we conclude that $T_{\rm kin}$ is down to 7.5$\pm{1}$ K in the center and
rises to $\sim$10--17 K in the outer region.  Because dust-gas coupling is efficient enough for $n_{\rm H_2}> 10^{4.5}$ cm$^{-3}$
\citep{Goldsmith01}, we can assume that $T_{\rm dust}=T_{\rm kin}$ in the core center based on our modeling. The central $T_{\rm
dust}$  is therefore  lower than \citet{Lippok13, Lippok16}'s 9.8--11.3 K, which explains our peak column density discrepancy as 
already noticed in other sources by \citet{Pagani15}.

\subsection{Cation abundance profiles}
\label{sec:discussion_iondepletion}

We find that \NtHp is significantly depleted at the center of L\,1512 ($n_{\rm H_2}=1.1\times 10^5$ cm$^{-3}$) and its abundance
starts to decrease at $n_{\rm H_2}$ of 3\,(east) to 8\,(west) $\times10^4$ cm$^{-3}$. This is one order of magnitude
lower than toward L\,183. By averaging the depletion factors calculated for the east and west sides, we obtain an averaged \NtHp
depletion factor of 27$^{+17}_{-13}$, which 
is larger than the factor of 6$^{+13}_{-3}$ in L\,183 previously found by \citet{Pagani07} 
while L\,183 even has a denser central $n_{\rm H_2}\geq 2 \times 10^6$ cm$^{-3}$. 
Strangely, \citet{Lee03} suggests that
\NtHp may be depleted in L\,1512 but not in another advanced and denser starless core, L\,1544 ($n_{\rm H_2}\geq 10^6$ cm$^{-3}$)
whereas \citet{Redaelli19} present new observations and analysis of L\,1544 for which they report a large volumetric depletion
factor ($>$ 100). 
For \NtDp, we find a depletion factor of 4$^{+2}_{-1}$, which is also higher than the factor of 2--2.5 found in
L\,183 \citep{Pagani07} but lower than the one reported for L\,1544 by \citet{Redaelli19}\footnote{It must be noted
that despite their claim, including in the title, Redaelli\,et\,al. were not the first to detect and explain \NtDp depletion.},
$\sim$15. It is surprising that the depletion factors do not correlate with the peak density among these three sources. One
of the possibilities is that the dust grain size in L\,1512 might be smaller than in the other cores keeping the depletion
timescale shorter \citep{Lee03}.  
Another possibility is the time it took to form the cores. 
L\,183 seems to have contracted more rapidly than the two other cores 
($\leq$0.7 Myr for L\,183, \citealt{Pagani13}; $\sim$1 Myr for L\,1544, \citealt{Redaelli19};
$\geq$1.4 Myr for L1512, see below Sect.\,\ref{sec:lifetime}), though this would have given more time for grain growth in L\,1512.
Many coupled dynamical and chemical models also cannot simulate such high depletion factor at $n_{\rm H_2}\simeq 10^5$ cm$^{-3}$ \citep[e.g.,][]{Aikawa01, Li02, Pagani13}, 
which might be caused by their too simplistic approach.
Consequently, more detailed modeling would be necessary to understand the physical and chemical evolution of L\,1512.

\begin{figure*}[t]
\sidecaption \includegraphics[width=12cm]{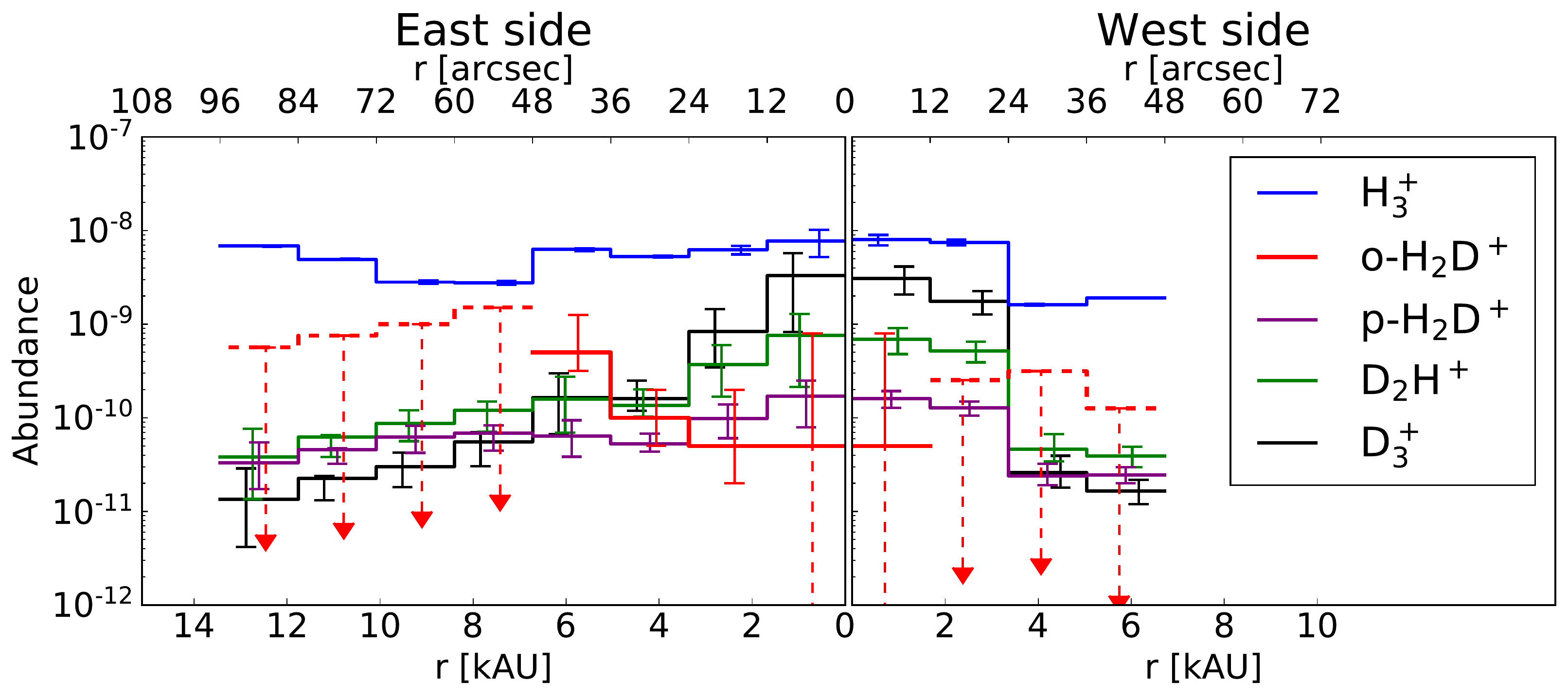} 
\caption \Hthp isotopolog abundances derived from the chemical
model constrained by the observations (in particular of ortho-\HtDp, which explains why we separately show the two spin states for
this species for the sake of clarity). 
\label{fig:modelresh3p}
\end{figure*}

From \NtHp and \NtDp, we can obtain the deuteration ratio, $X$(\NtDp)/$X$(\NtHp), which is increasing from an upper limit of
$\leq$0.05 at large radii to the maximum of 0.34$^{+0.24}_{-0.15}$ at the center of L\,1512.
The higher deuteration ratio at the center seems contradictory with the decreasing abundance of ortho-\HtDp toward the center
(by a factor of $\lesssim$10).  It can be explained if the ortho-\HtDp depletion is due to the further deuteration of \Hthp.
Namely,  \HtDp is converted to  \DtHp and \Dthp, which is supported by our chemical modeling best-fit result,
$X$(\Dthp)/$X$(ortho-\HtDp)\,$\approx$\,64 and $X$(\DtHp)/$X$(ortho-\HtDp)\,$\approx$\,15 at the innermost layer (Fig.\,\ref{fig:modelresh3p}). The contrast
is higher than toward L\,183 \citep{Pagani09b, Pagani13} 
where \DtHp is comparable to \HtDp in abundance and \Dthp less than a
factor of 10 higher than \HtDp. Moreover, in L\,183, the ortho-\HtDp does not show depletion toward the center. This can be linked to the longer
timescale of evolution of L\,1512 as discussed in \citet{Pagani13} appendix B, where it is shown that the para-\DtHp/ortho-\HtDp is much
higher in the slow collapsing case than in the fast collapsing case (the reader should be aware that para-\DtHp is only a small fraction of total \DtHp
when comparing the two papers).

Remaining gas-phase CO can react with deuterated \Hthp and form \DCOp \citep{Pagani11}, we find that \DCOp happens to deplete at
$n_{\rm H_2}$ of 3\,(east) to 7\,(west) $\times 10^4$ cm$^{-3}$, and reaches a depletion factor of 9$^{+21}_{-3}$ at the center, which is
comparable to the factor of $\geq 17$ found in L\,183 \citep{Pagani12}. Like in L\,183, \DCOp depletes because its parent
molecule, CO, freezes out onto dust grains much more than what the growth of \Hthp deuteration can compensate for
\citep{Pagani12}.  Likewise, \NtHp and \NtDp deplete because of the freeze-out of N$_2$.  CO and N$_2$ depletion are
discussed hereunder.

\subsection{CO and N$_2$ depletion}
\label{sec:discussion_depletion}

It is impossible to measure $X$(CO) and $X$(N$_2$) directly in a dense core. 
This is because the emission of the low-$J$ transitions
of CO (including its isotopologs) usually becomes optically thick even for rarer isotopologs such as \CetO and C$^{17}$O,
for which the emission is dominated by the envelope and N$_2$ has no transition detectable\footnote{The lowest
N$_2$ rotational transition $J$=2--0 at 357.808 GHz is a very weak electrical quadrupolar transition, and in any case
would be blocked by the telluric opacity if it were detectable.} in starless cores
\citep{Pagani12}.

In Sect.\,\ref{sec:analysis_chem}, we presented a chemical model to relate the four observed cations (\NtHp, \NtDp, \DCOp,
and ortho-\HtDp) to CO and N$_2$ in the depleted region to retrieve their abundance profiles. Figure\,\ref{fig:chem_model_CO_N2} shows the
abundance profiles of \twvCO, \CetO, and N$_2$. 
We find that N$_2$ has a depletion factor of $\sim$200 compared with its
maximum values of $3\times 10^{-5}$.
For CO isotopologs, \twvCO has a depletion factor of $\sim$220 compared with its
maximum values of $5\times 10^{-5}$. 
Compared with the standard \twvCO abundance of 1--$2\times 10^{-4}$ \citep{Pineda10}, \twvCO has a depletion factor of $\sim$430--870.
\CetO would also have a depletion factor of $\sim$220 compared with its maximum value of $1\times 10^{-7}$ in the outer region since we fix the $^{12}$C$^{16}$O/$^{12}$\CetO ratio of 500, and a factor\footnote{The discrepancy between the two depletion factors, 430--870 of \twvCO and 370 of \CetO, is because the standard \twvCO abundance of 1--$2\times 10^{-4}$ and \CetO abundance of $1.7\times 10^{-7}$ are not compatible with a standard \twvCO/\CetO ratio of 500.} of $\sim$370
compared to the standard \CetO abundance of $1.7\times 10^{-7}$ \citep{Frerking82}, 
which are much larger than the factors of 30 and 25 found by \citet{Lippok13}\footnote{However, we could not
reproduce the \citet{Lippok13} model results (the emergent intensities being typically a factor of 2 too low based on their values) and the discrepancy
has not been identified, but could be linked to a possible confusion between H and \Ht densities in the \citet{Lippok13} paper.} and \citet{Lee03}, respectively. Although both studies try to recover volumetric abundances of \CetO and not
simply column-density averages, the abundance drop they derive can only be a lower limit because the line from the center is too
weak to constrain the models. A factor of 30 already indicates that the contribution of the central part to the total intensity is
less than $\sim$10\%, which would require a very high signal-to-noise ratio to evaluate and still does not guarantee that the measured
variation is not simply due to irregularities in the cloud shape rather than abundance variations  (see \citealt{Pagani10b} for a
similar case). 

The abundance of CO and N$_2$ in starless cores is a matter of long debate (see, e.g., \citealt{Pagani12,Nguyen18})
since the daughter species of N$_2$, \NtHp, is present at much higher densities than CO itself in starless cores, while CO and
N$_2$ have identical masses and similar sticking coefficients \citep{Bisschop06} and their freeze-out rates are thus comparable.
For L\,1512, \twvCO and N$_2$ profiles are similar to each other within their error bars. This is different from the results
reported by \citet{Pagani12} who found a different N$_2$ profile in L\,183, where $X$(N$_2$) is less than $X$(\twvCO) by about two
orders of magnitude at the edge of the starless core. These authors suggest that N$_2$ might  still be forming from atomic N in
L\,183. In that case, the depletion and production rates of N$_2$ would partly compensate each other. In contrast with
L\,183, we find that L\,1512 has chemically evolved on a much longer timescale so that N$_2$ chemical production from atomic N has
reached steady state. This could explain the similar depletion profiles. The fact that the CO depletion in L\,183 is $\sim$3 times deeper than in L\,1512 is mostly due to its peak density being 20 times higher, making collisions with dust grains much more frequent, but the difference is partly compensated by a much longer evolution time in L\,1512.
The L1512 case is interesting because, by showing that N$_2$ and CO profiles are identical while \NtHp is tracing the core and not CO, it demonstrates that the N$_2$/CO problem is not correctly formulated. \NtHp cannot be directly compared to
CO because it is a daughter species of N$_2$ and its abundance therefore also depends on the abundance of \Hthp. We would need to
compare HCO$^+$ to \NtHp, which is not possible in a simple way. Also, \NtHp is a trace species, typically at the 10$^{-10}$
abundance level, at least two orders of magnitude less abundant than CO even depleted by a factor 2000 as in L\,183. A correct
approach is therefore to compare DCO$^+$ and \NtDp; in most clouds, it is clearly visible that DCO$^+$ is more abundant
than \NtDp\ because its emission is usually as extended as that of \NtHp itself. There is therefore no N$_2$/CO crisis. However,
tracing the abundance of both species in starless cores remains important since it helps the understanding of the chemical evolution of
the core.

\subsection{Lifetime scale}
\label{sec:lifetime}
Figure\,\ref{fig:chem_model} shows that the chemical solution of each layer meets the observed abundances within $\sim$0.2--2.6 Myr.
Since the physical structure is kept constant in our pseudo time-dependent chemical modeling, we see that the chemical process is
relatively accelerated in the inner layers compared with the outer layers. Therefore, we underestimate the timescale for the
inner layers, which were not very dense at the beginning of contraction. 
In addition, only the deuteration ratio of \NtDp/\NtHp and the abundance of ortho-\HtDp are used to constrain the chemical age, while $X$(\DCOp) is used to determine the CO abundance.
It follows that our result gives a lower limit on the
lifetime scale of L\,1512 ($\sim$1.43 Myr) that is set by the third western layer, 
which is the last one in our chemical model to reach the observed \NtDp/\NtHp ratio.
As a lower limit is required, this value is set at the beginning of the time range of the third western layer as shown on the top right panel (the chemical solutions for the western deuteration ratios) in Fig.\,\ref{fig:chem_model}.
We also note that since the third western layer has no ortho-\HtDp detection but only an upper limit on its abundance, the time limit for this layer is the time it takes to reach $X$(\NtDp)/$X$(\NtHp) under the assumption of $X$(ortho-\HtDp) at the maximum of its possible range, that is the fastest time of that layer.
Therefore, L\,1512 is presumably older than 1.43 Myr.
We might ask whether the initial OPR of 3 is warranted. If we set an initial OPR as low as 0.01, we cannot find a good fit for both sides.
On the other hand, if we use an OPR of 0.5 as the initial value, as measured in the diffuse medium \citep{Crabtree11} toward a few lines of sight, we can only find very marginal fits for both sides and the time remains somewhat above 0.5 Myr. This suggests that the OPR cannot be lower than $\sim$1.

\citet{Pagani09b} applied the same pseudo time-dependent chemical modeling on L\,183 and found the lower limit on L\,183 lifetime scale to be 0.15--0.2 Myr. Later, \citet{Pagani13} conclude that the contraction of L\,183 follows the free-fall timescale ($\sim$0.5--0.7 Myr). 
Namely, L\,1512 is older than the L\,183  starless core at least by a factor of $\sim$2--3. This is in agreement with our conclusion from  Sect.\,\ref{sec:discussion_depletion} that L\,1512 may be chemically more evolved than L\,183, while physically L\,183 has reached a higher density in a shorter time.  Consequently, ambipolar
diffusion may have slowed down the contraction of L\,1512 or even halted it to the present state, while it has had no impact
on L\,183. This would imply that  the magnetic field is stronger in L\,1512 than in L\,183. This will be the subject of a
follow-up study of L\,1512.

\section{Conclusions}\label{sec:conclusion}

We established the extinction map of L\,1512 from deep $JHK_s$ and Spitzer/IRAC data.
We performed a non-LTE radiative transfer calculation with an onion-shell model to 
reproduce the observed spectra of L1512. 
We obtained separately, for the east and west sides, 1D spherically symmetric profiles of the physical parameters and chemical abundances of \NtHp, \NtDp, \DCOp, and ortho-\HtDp.
Afterward, we
used a time-dependent chemical model to estimate the lower limit on the lifetime scale of L\,1512 by fitting the chemical solutions
to the observed abundances. Our chemical model captures the main reactions in the \Hthp\ deuteration fractionation and assumes that
heavy species are totally depleted except for CO and N$_2$, which are partially depleted. Thus, by relating the observed abundances to CO and N$_2$ in the depleted
region, we could also obtain the CO and N$_2$ profiles and their depletion factors. We summarize our results as follows:
\begin{enumerate}
\item We derived the central molecular hydrogen density of $1.1\times10^5$ cm$^{-3}$ from the dust extinction measured from NIR 
and MIR maps and the central kinetic temperature of 7.5$\pm$1 K from \NtHp observations.
At such density, gas is thermalized with dust.
\item The depletion factors of \NtHp, \NtDp, \DCOp, and ortho-\HtDp are 27$^{+17}_{-13}$, 4$^{+2}_{-1}$, 9$^{+21}_{-3}$, and $\lesssim$10.
Compared with \NtHp, the smaller \NtDp depletion factor is due to the deuterium fractionation enhanced from a upper limit 
of $\leq 0.05$ at large radii to 0.34$^{+0.24}_{-0.15}$ in the center.
The depletion of ortho-\HtDp suggests that ortho-\HtDp is further deuterated to \DtHp and \Dthp.
\item The depletion of \NtHp and \NtDp traces the freezing out of N$_2$ toward the center of L\,1512.
We find that N$_2$ has a depletion factor of $\sim$200 compared to its maximum abundance of $4.7\times 10^{-5}$.
Likewise, the depletion of \DCOp indicates the freeze-out of CO.
We find that \twvCO and \CetO have a depletion factor of $\sim$220 between internal and external layers. If we compare their minimum abundance to their standard (literature) abundance, we find the depletion factor is about 400 for both isotopologs.
\item The similarity of CO and N$_2$ profiles suggests that L\,1512 may have chemically evolved long enough and in particular that N$_2$ chemistry has reached its steady state.
This could explain that L\,1512 has a higher depletion factor of N$_2$ compared to the  denser starless core, L\,183.
\item L\,1512 is presumably older than 1.4 Myr.
We conclude that ambipolar diffusion is the dominant core formation mechanism for this source.
\end{enumerate}
In summary, our results present a precise description of density, temperature, and molecular abundances in a starless core.

\begin{acknowledgements}
S.J.L. and S.P.L. acknowledge the support from the Ministry of Science and Technology (MOST) of Taiwan with grant MOST 106-2119-M-007-021-MY3 and MOST 106-2911-I-007-504.
Nawfel Bouflous and Patrick Hudelot (TERAPIX data center, IAP, Paris, France) are warmly thanked for their help in
preparing the CFHT/WIRCAM observation scenario and for performing the data reduction.
This work was supported by the
Programme National “Physique et Chimie du Milieu Interstellaire” (PCMI)
of CNRS/INSU with INC/INP co-funded by CEA and CNES and by Action F\'ed\'eratrice Astrochimie de l'Observatoire de Paris.
This work is based in part on observations carried out under project numbers 152-13, 039-14, 112-15, and
D01-17 with the IRAM 30m telescope. IRAM is supported by INSU/CNRS (France), MPG (Germany) and IGN
(Spain). The James Clerk Maxwell Telescope is operated by the East Asian Observatory on behalf of The
National Astronomical Observatory of Japan, Academia Sinica Institute of Astronomy and Astrophysics,
the Korea Astronomy and Space Science Institute, the National Astronomical Observatories of China and
the Chinese Academy of Sciences (Grant No. XDB09000000), with additional funding support from the
Science and Technology Facilities Council of the United Kingdom and participating universities in the
United Kingdom and Canada, the JCMT data were collected during programs ID M13BC01 (SCUBA-II),
M15BI046 (HARPS, and M17BP043 but data were lost due to an incorrect tuning). This research made use
of the Aladin interface, the SIMBAD database, operated at CDS, Strasbourg, France, and the VizieR
catalog access tool, CDS, Strasbourg, France. This research has made use of the NASA/IPAC Infrared Science Archive, which is operated by the Jet Propulsion Laboratory, California Institute of Technology, under contract with the National Aeronautics and Space Administration. This work is based in part on observations made with the Spitzer Space Telescope, which is operated by the Jet Propulsion Laboratory, California Institute of Technology under a contract with NASA. This research used the facilities of the Canadian Astronomy Data Centre operated by the National Research Council of Canada with  the support of the Canadian Space Agency.
Green Bank Observatory is a facility of the National Science Foundation and is operated by Associated Universities, Inc. 
\end{acknowledgements}

\bibliographystyle{aa}
\bibliography{ref}

\begin{appendix}

\section{Spectral line observations}\label{app:all_spectra}

We present spectral line observations other than the main horizontal and vertical cuts and the full \CetO horizontal cut.

\begin{figure*}[h!]
	\centering
	\includegraphics[width=15cm]{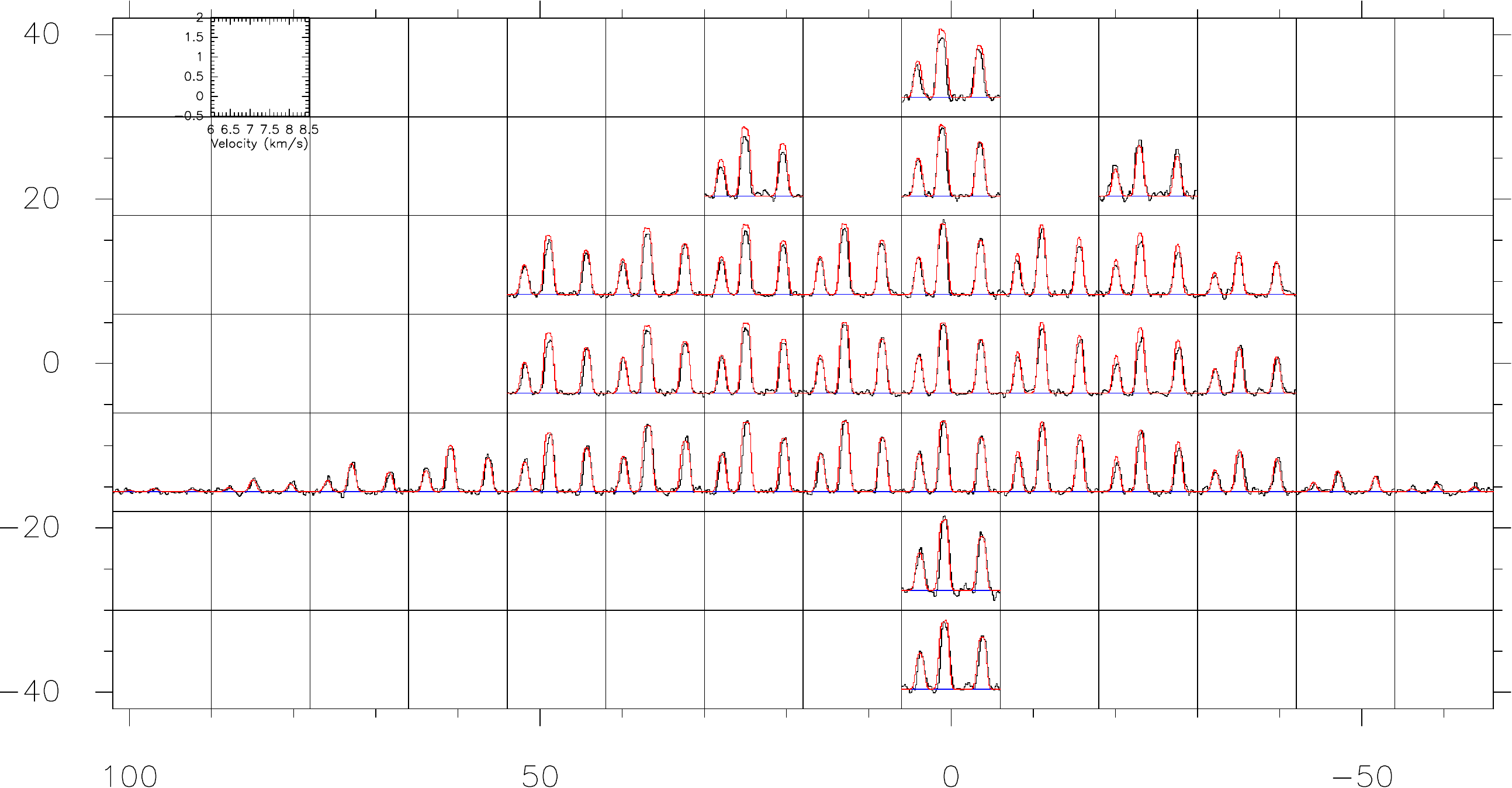}
	\caption{
    \NtHp (1--0) spectra.
    The $x$- and $y$-axes of the grid are the RA and Dec offsets with respect to the center of L\,1512.
    Each cell shows the observational spectra as black, the modeled spectra as red, and the baselines as blue. The dimension of $T_{\rm A}^*$ and $V_{\rm LSR}$ at each cell are denoted in one of the empty cells. Only the central triplet is shown for clarity.
    \label{fig:n2hp10_spectra}
    }
\end{figure*}

\begin{figure*}[h!]
	\centering
	\includegraphics[width=14cm]{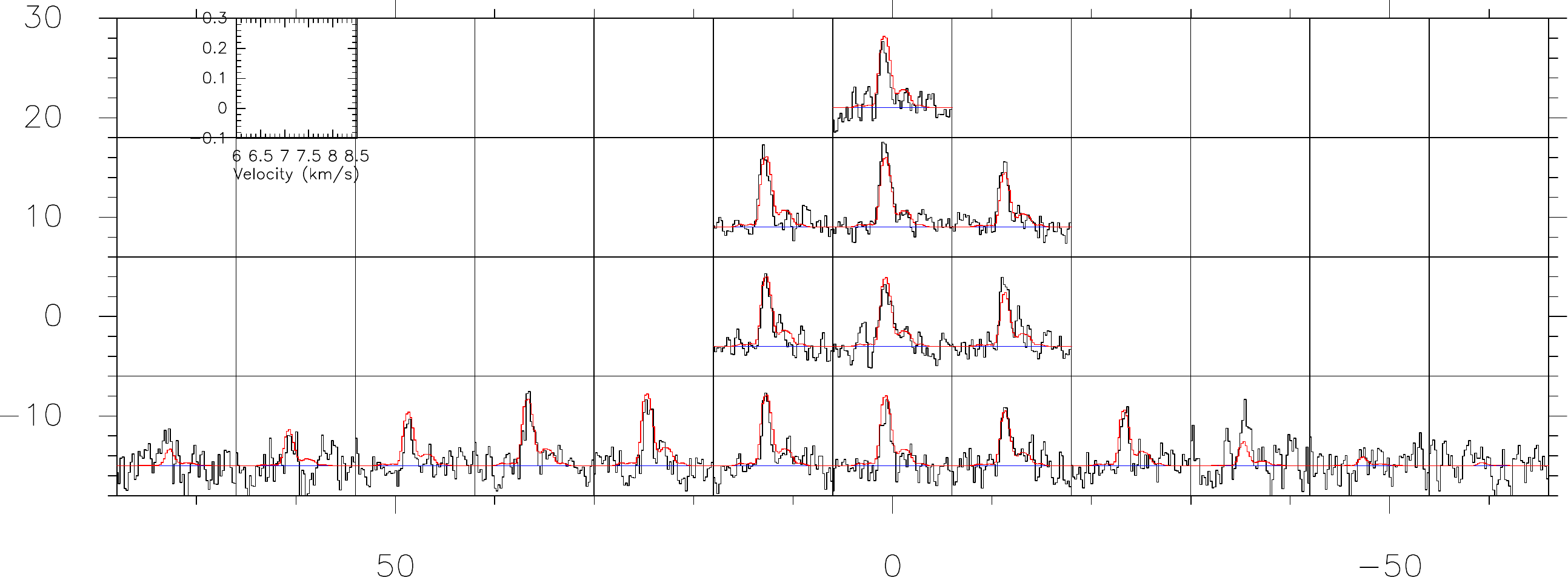}
	\caption{
    \NtHp (3--2) spectra.
    Annotations are the same as in Fig.\,\ref{fig:n2hp10_spectra}.
    \label{fig:n2hp32_spectra}
    }
\end{figure*}

\begin{figure*}[h!]
	\centering
	\includegraphics[width=9.5cm]{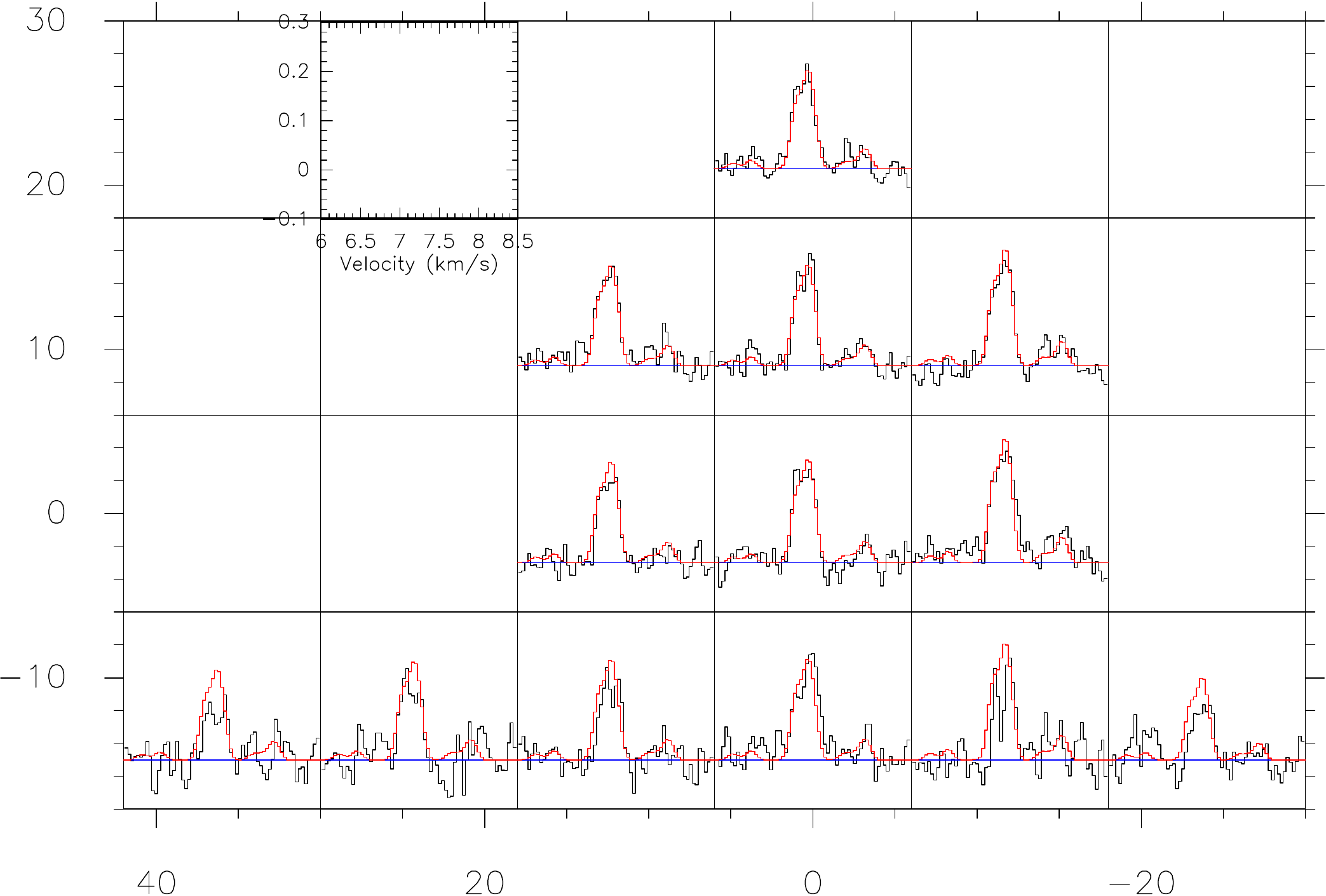}
	\caption{ 
    \NtDp (2--1) spectra.
    Annotations are the same as in Fig.\,\ref{fig:n2hp10_spectra}.
    \label{fig:n2dp21_spectra}
    }
\end{figure*}

\begin{figure*}[h!]
	\centering
	\includegraphics[width=15cm]{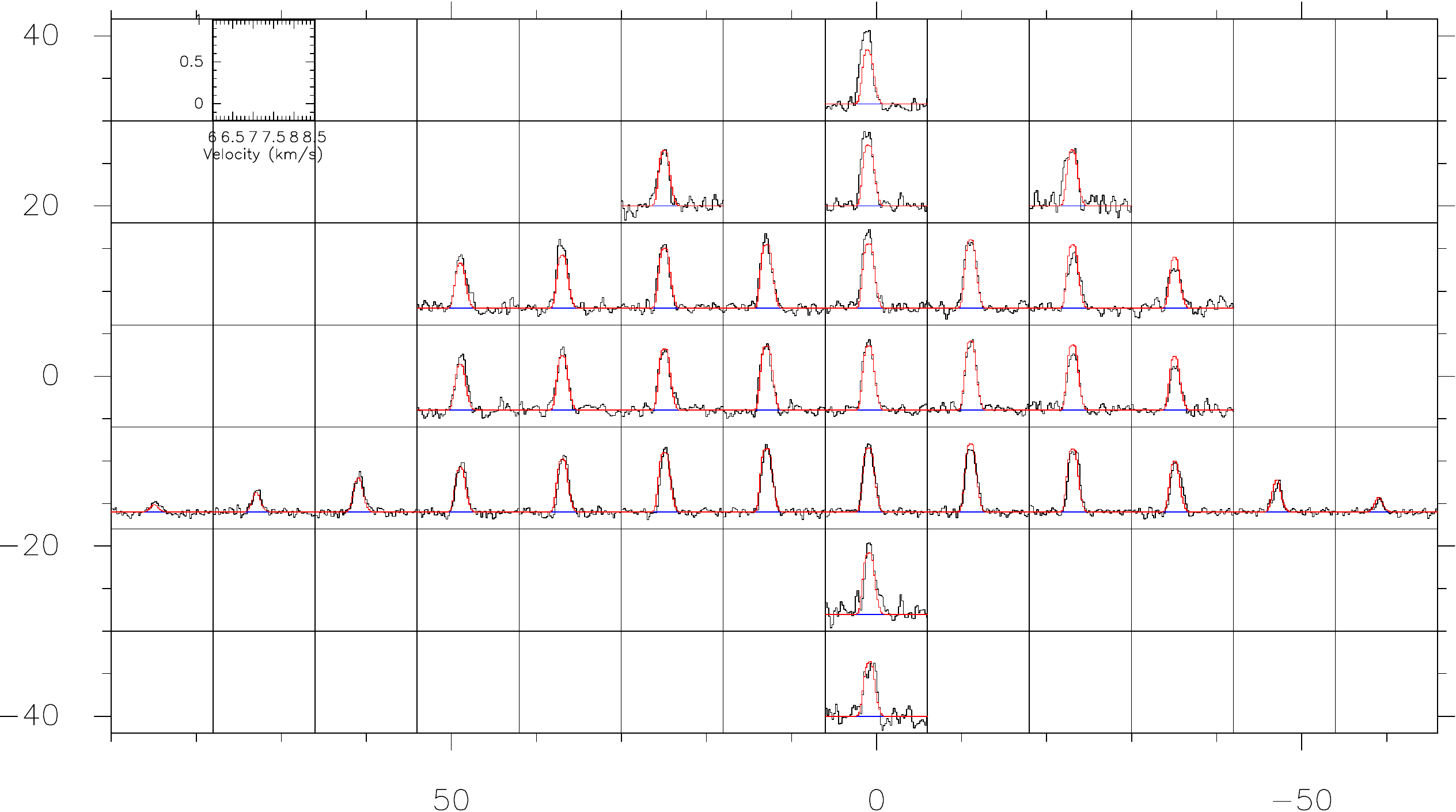}
	\caption{
    \DCOp (2--1) spectra.
    Annotations are the same as in Fig.\,\ref{fig:n2hp10_spectra}.
    \label{fig:dcop21_spectra}
    }
\end{figure*}

\begin{figure*}[h!]
	\centering
	\includegraphics[width=12cm]{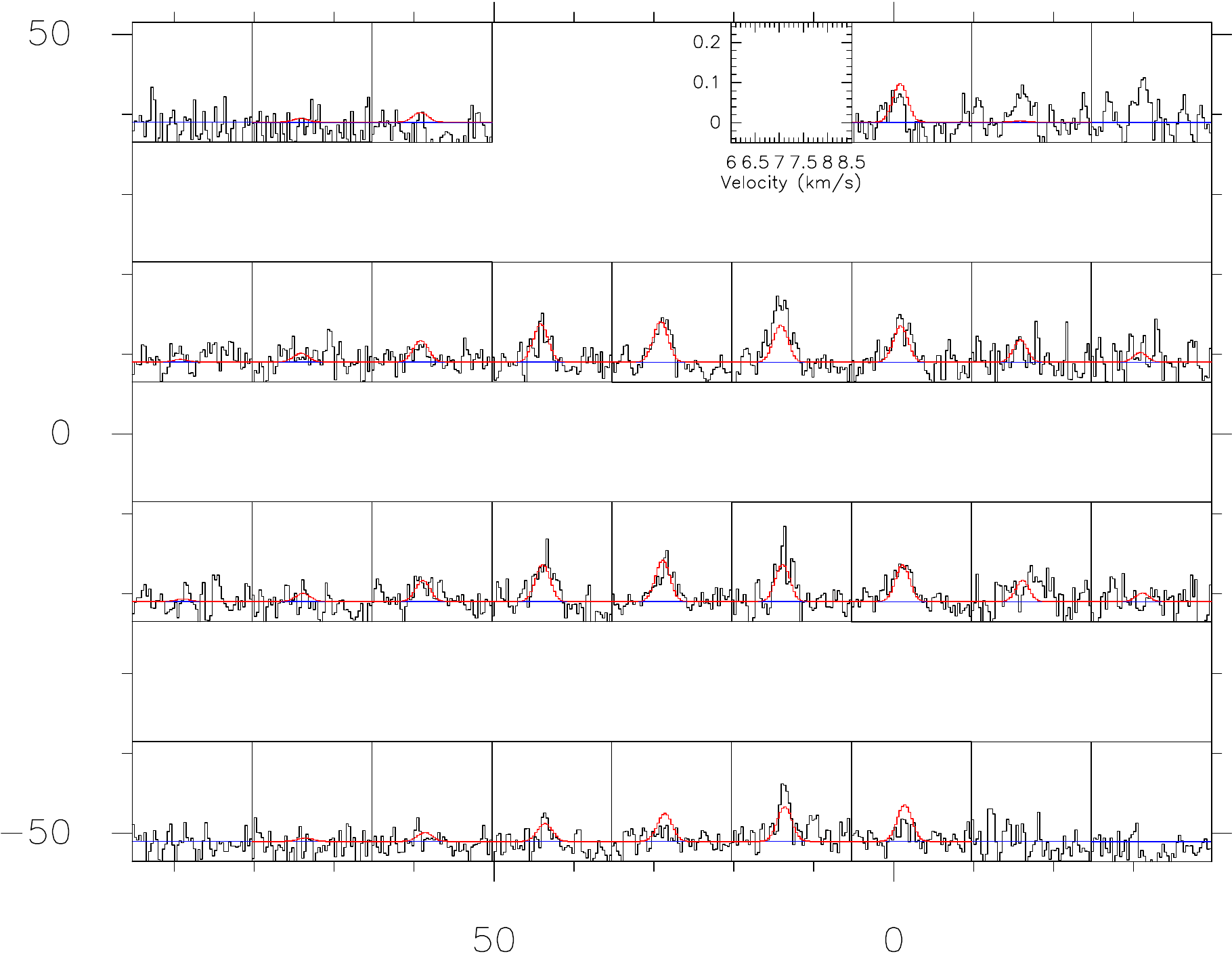}
	\caption{
    \HtDp (\oHtDpgrd) spectra.
    Annotations are the same as in Fig.\,\ref{fig:n2hp10_spectra}.
    \label{fig:h2dp_spectra}
    }
\end{figure*}

\begin{figure*}[h!]
	\centering
	\includegraphics[width=19cm]{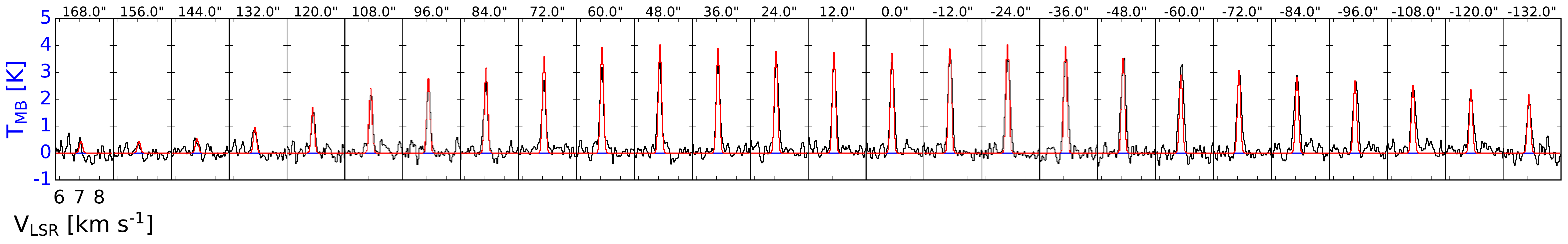}
	\caption{
    \CetO (1--0) spectra along the main horizontal cut at $\Delta$Dec=$-$12\arcsec.
    Annotations are the same as in Fig.\,\ref{fig:n2hp10_spectra} except that the emissions are in the $T_{\rm MB}$ scale.
    \label{fig:c18o10_spectra}
    }
\end{figure*}

Figures \ref{fig:n2hp10_spectra}, \ref{fig:n2hp32_spectra}, \ref{fig:n2dp21_spectra}, \ref{fig:dcop21_spectra}, and \ref{fig:h2dp_spectra} show all the single pointing observations in black lines and models of \NtHp (1--0), \NtHp (3--2), \NtDp (2--1), \DCOp (2--1), and ortho-\HtDp (\oHtDpgrd) in red lines, respectively. 
The models are reproduced with our dissymmetric onion-shell model (Sect.\,\ref{sec:analysis} and Table \ref{tab:rad_model}).
The full \CetO (1--0) spectra along the main horizontal cut is shown in Fig.\,\ref{fig:c18o10_spectra}, which is an extension of the \CetO spectra in Fig.\,\ref{fig:rad_model}. 
Since the \CetO emission extending toward the outskirts becomes asymmetric to the core center, we adopt two slab models for reproducing the spectra at $\Delta$Dec $\leq -72$\arcsec (west) and $\Delta$Dec $\geq 120$\arcsec (east), respectively, instead of the above onion-shell model.
To fit the observations in the outskirts, the \CetO abundance of the slab models are fixed to 10$^{-7}$ (Sect.\,\ref{sec:analysis_chem}) and $N_{\rm H_2}$ is found to be 3.5--5.6$\times 10^{21}$ cm$^{-2}$ for the western slab and 0.8--2.6$\times 10^{21}$ cm$^{-2}$ for the eastern slab.

\section{N$_{2}$D$^{+}$--H$_2$  collisional rate coefficients}
\label{N2D+collcoef}

Rate coefficients for the N$_{2}$H$^{+}$--H$_2(J=0)$ collisional system have been provided by \cite{Lique14}.
Hyperfine-structure-resolved rate coefficients, based on a highly accurate ab initio potential energy surface (PES) \citep{Spielfiedel15}, were calculated for temperatures 
ranging from 5 to 70 K. The new rate coefficients were found to be significantly different from  the N$_2$H$^+$--He rate coefficients previously published \citep{Daniel05}. 

Recent studies \citep{Dumouchel12,Flower15,Dumouchel17} have shown that isotopic effects in inelastic collisions can be important, especially for H/D substitution. Hence, we 
decided to compute actual N$_{2}$D$^{+}$--H$_2(J=0)$ rate coefficients.

Within the Born-Oppenheimer approximation, the full electronic ground state potential is identical for N$_{2}$H$^{+}$--H$_2$ and N$_{2}$D$^{+}$--H$_2(J=0)$ systems and 
depends only on the mutual distances of the five atoms involved. Then, we used for the scattering calculations the N$_{2}$H$^{+}$--H$_2$ PES of \cite{Spielfiedel15} and the 
``adiabatic-hindered-rotor'' treatment, which allows para-H$_2(J=0)$ to be treated as if it were spherical. 
The major difference between the N$_{2}$H$^{+}$--H$_2$ and N$_{2}$D$^{+}$--H$_2(J=0)$ PESs is the position of the center of mass taken for the origin of the Jacobi coordinates. 
For N$_{2}$D$^{+}$--H$_2(J=0)$ calculations, we considered the effect of the displacement of the center of mass.

Since the nitrogen atoms possess a non-zero nuclear spin ($I=1$), the N$_2$D$^+$ rotational energy levels, such as those of N$_{2}$H$^{+}$, are split in hyperfine levels, which are 
characterized by the quantum numbers $J$, $F_1$ and $F$. In this case, $F_1$
results from the coupling of the rotational level $\vec{J}$ with $\vec{I_1}$ ($\vec{F_1} =
\vec{J} + \vec{I_1}$, $I_1$ being the nuclear spin of the first nitrogen atom) and $F$ results from the coupling of
$\vec{F_1}$ with $\vec{I_2}$ ($\vec{F} = \vec{F_1} + \vec{I_2}$, $I_2$ being the nuclear spin of the second nitrogen atom).
The D atom also possesses a non-zero nuclear spin. However, in the astronomical observations, the hyperfine structure due to D is not resolved and is then neglected in the 
calculations.

The hyperfine splitting of the N$_2$D$^+$ energy levels is very small. Considering that the hyperfine levels are degenerate, we
simplified the hyperfine scattering problem using recoupling techniques \citep{Faure12}. Then, we performed Close-Coupling
calculations \citep{Arthurs60} of the pure rotational excitation cross sections (neglecting the hyperfine structure) using the
MOLSCAT program \citep{molscat94}. The N$_2$D$^+$ energy levels were computed using the rotational constants of \cite{Dore09}.
Calculations were carried out for total energies up to 500~cm$^{-1}$. Parameters of the integrator were tested and adjusted to
ensure a typical precision to within 0.05 \AA$^2$ for the inelastic cross sections. At each energy, channels with $J$ up to 28
were included in the rotational basis to converge the calculations for all the transitions including N$_2$D$^+$ levels up to $J =
7$. Using the recoupling technique, the hyperfine state-resolved cross sections were obtained for all hyperfine levels up to
$J=7$.

From the calculated cross sections, we can obtain the corresponding thermal rate coefficients at temperature $T$ by an average over the collision energy ($E_{\rm c}$), i.e.,

\begin{eqnarray}
\label{thermal_average}
k_{\rm \alpha \rightarrow \beta}(T) & = & \left(\frac{8}{\pi\mu k^3_{\rm B} T^3}\right)^{\frac{1}{2}}  \nonumber\\
&  & \times  \int_{0}^{\infty} \sigma_{\rm \alpha \rightarrow \beta}\, E_{\rm c}\, e^{-\frac{E_{\rm c}}{k_{\rm B}T}}\, dE_{\rm c},
\end{eqnarray}
where $\sigma_{\rm \alpha \to \beta}$ is the cross section from initial level $\alpha$ to final level $\beta$, $\mu$ is the reduced mass of the system, and $k_{\rm B}$ is Boltzmann's 
constant. 

Using the computational scheme described above, we obtained N$_2$D$^+$--H$_2(J=0)$ rate coefficients for temperatures up to 70 K. The complete set of (de-)excitation rate 
coefficients with 
$J,\,J'\le 7$ will be made available through the LAMDA \citep{Schoier05} and BASECOL \citep{Dubernet13} databases.

\begin{figure*}[h!t]
\centering
\includegraphics[width=8.0cm]{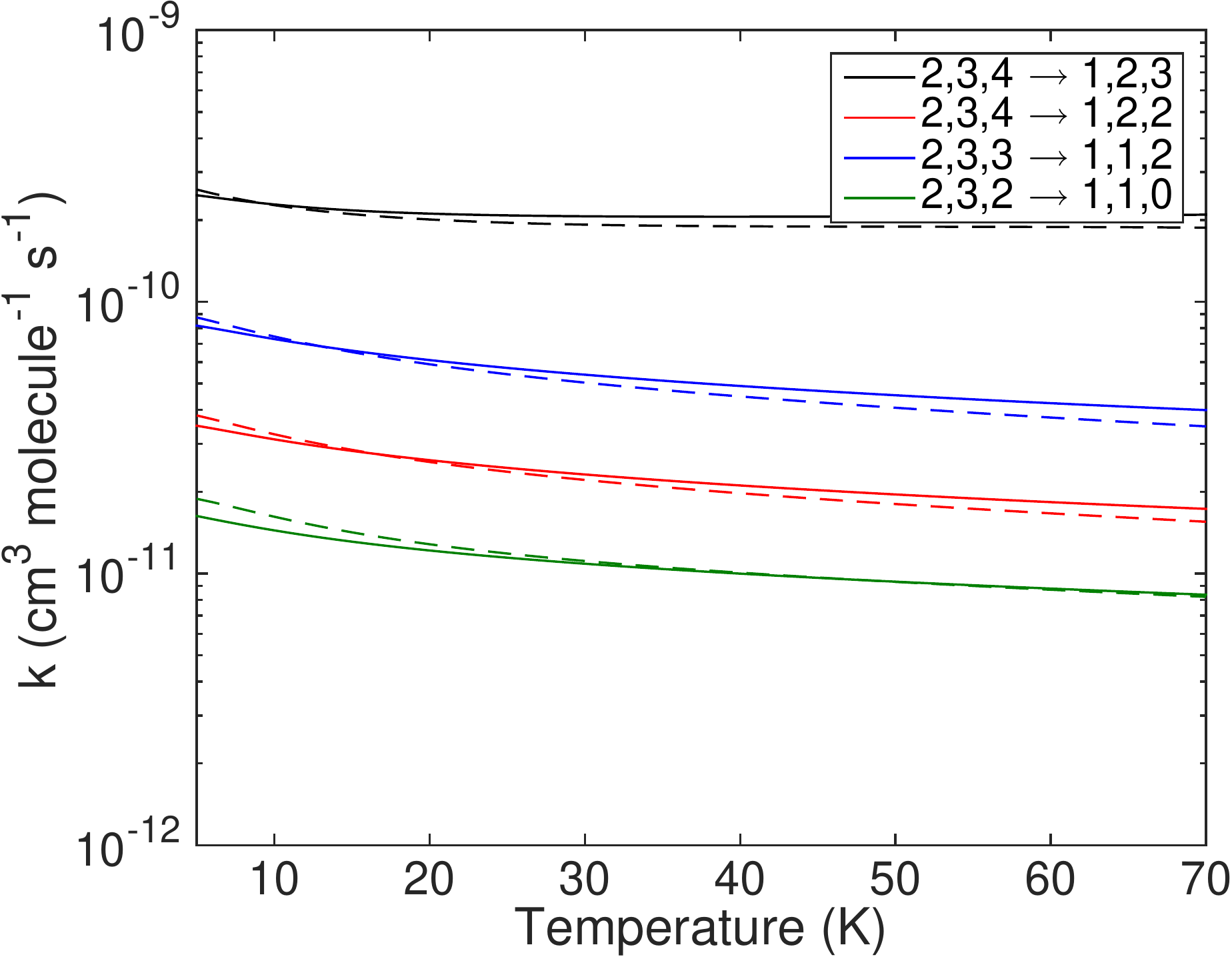}
\includegraphics[width=8.0cm]{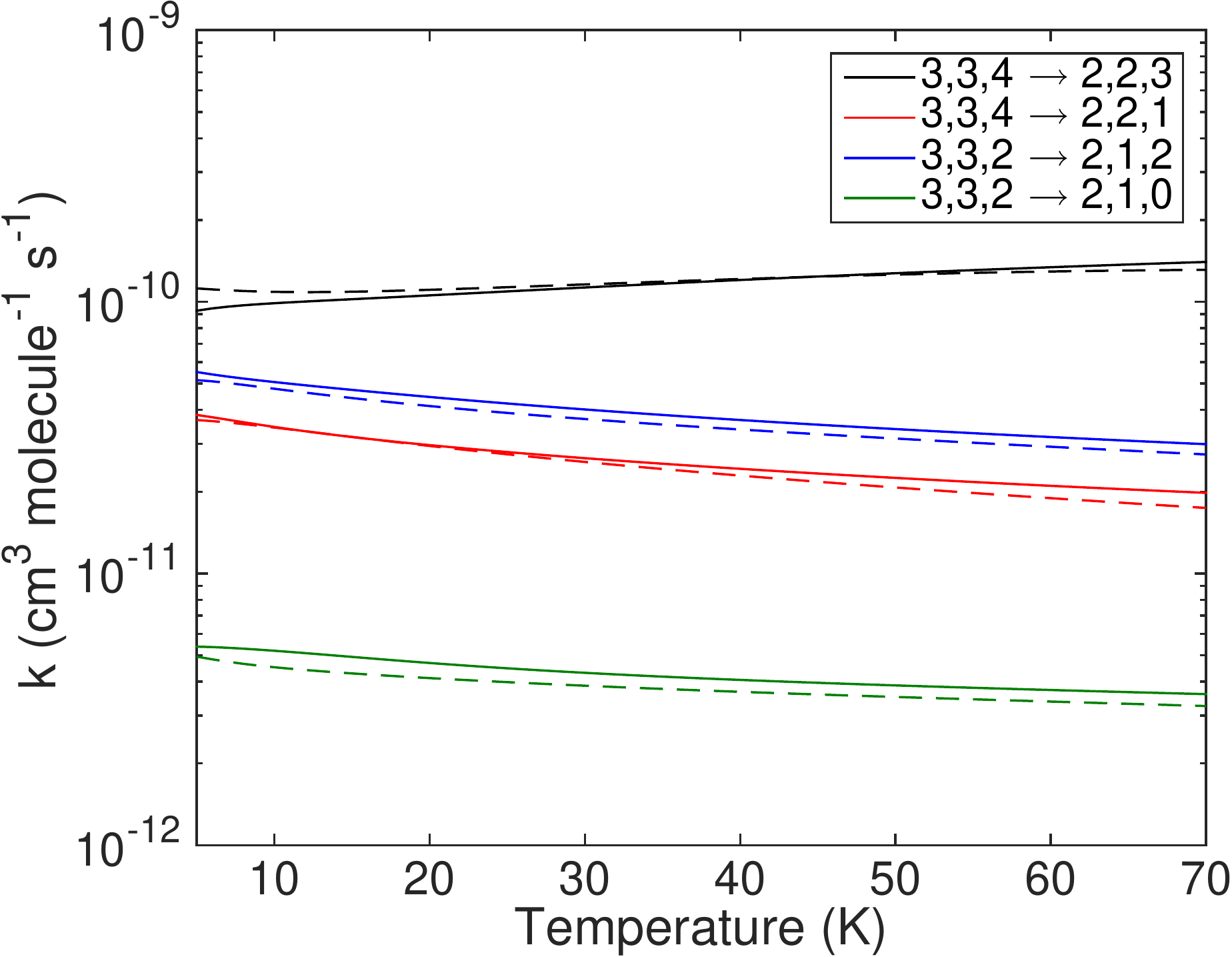}
\caption{Temperature variation of the hyperfine-resolved
N$_2$H$^+$--H$_2(J=0)$ (solid lines) and N$_{2}$D$^{+}$--H$_2(J=0)$ (dotted lines) rate coefficients for $J=2,F \to J'=1,F'$ and $J=3,F \to J'=2,F'$
transitions.} \label{fig:rates}
\end{figure*}

Figure \ref{fig:rates} presents the temperature variation of the
N$_2$H$^+$--H$_2(J=0)$ and N$_2$D$^+$--H$_2(J=0)$ rate coefficients for selected $J=2,F_1,F \to
J'=1,F_1',F'$ and $J=3,F_1,F \to
J'=2,F_1',F'$ transitions. 

As is shown, a very good agreement is found between N$_2$H$^+$--H$_2(J=0)$ and N$_{2}$D$^{+}$--H$_2(J=0)$ rate coefficients.
Both sets of data agree within a few percents over all the temperature range. The largest deviations are seen at a low temperature ($T
< 10$~K) that characterizes cold molecular clouds. The N$_2$H$^+$ over N$_2$D$^+$ rate coefficients ratio depends on the
temperature showing that extrapolation techniques are not suited for the estimation of N$_{2}$D$^{+}$ collisional data. The
differences are due to both the center-of-mass shift on the interaction potential and the use of isotopolog specific energies of
the levels.

\section{Best-fit physical and abundance profiles}\label{app:rad_model_table}

We present the best-fit quantities with their error ranges for each layer in the onion-shell model from Fig.\,\ref{fig:rad_model} and Fig.\,\ref{fig:chem_model_CO_N2} in Table\,\ref{tab:rad_model}.

\begin{sidewaystable*}[hb]
    \caption{Onion-shell model parameters.}
    \centering
    \begin{tabular}{ccrccrccccc}
    \hline\hline
    $r$        & $n_{\rm H_2}$& $T_{\rm kin}$ & $X$(\NtHp)                 & $X$(\NtDp)   & $D_{\rm N_2H^+}$$^{\tablefootmark{a}}$ & $X$(\DCOp)                 & $X$(o-\HtDp) & $X$(CO)      & $X$(C$^{18}$O) & $X$(N$_2$) \\ 
    (kAU)      & (cm$^{-3}$)  & (K)           &                            &              &                                        &                            &              &              &                &   \\
    \hline
    \noalign{\smallskip}
    \multicolumn{11}{c}{East side}\\
    \noalign{\smallskip}
    \hline
    \noalign{\smallskip}
    0.00--1.68& 1.06(5)$^{\tablefootmark{b}}$ & 7.5$_{-2}^{+4}$ & 3.31$_{-2.26}^{+3.30}$(-11) & 1.13$_{-0.42}^{+1.13}$(-11) & 0.34$_{-0.27}^{+0.48}$ & 1.57$_{-0.32}^{+6.31}$(-11) & 5.00$_{-\inf}^{+74.24}$(-11) & 3.16$_{-2.16}^{+6.84}$(-7) & 6.32(-10) & 1.45$_{-0.84}^{+2.03}$(-7) \\
    \noalign{\smallskip}
    1.68--3.36   & 8.43(4) & 8.2$_{-2}^{+1}$  & 9.20$_{-3.40}^{+2.38}$(-11) & 1.13$_{-0.42}^{+1.13}$(-11) & 0.12$_{-0.06}^{+0.13}$ & 6.40$_{-1.32}^{+1.66}$(-11) & 5.00$_{-3.00}^{+14.91}$(-11) & 5.19$_{-2.21}^{+3.83}$(-6) & 1.04$_{-0.38}^{+2.25}$(-8) & 3.77$_{-0.91}^{+1.20}$(-7) \\
    \noalign{\smallskip}
    3.36--5.04   & 6.00(4) & 8.5$_{-2}^{+1}$  & 4.36$_{-0.90}^{+1.13}$(-10) & 2.20$_{-0.45}^{+0.57}$(-11) & 0.05$_{-0.01}^{+0.02}$ & 8.72$_{-1.79}^{+2.26}$(-11) & 1.00$_{-0.50}^{+0.99}$(-10) & 1.00$_{-0.50}^{+0.99}$(-5) & 2.00$_{-0.74}^{+1.17}$(-8) & 3.51$_{-1.44}^{+2.45}$(-6) \\
    \noalign{\smallskip}
    5.04--6.72   & 4.21(4) & 9.2$_{-4}^{+1}$  & 9.06$_{-5.46}^{+2.35}$(-10) & 5.01$_{-1.03}^{+1.30}$(-11) & 0.06$_{-0.04}^{+0.02}$ & 1.03$_{-0.21}^{+0.27}$(-10) & 5.00$_{-1.85}^{+7.56}$(-10) & 9.16$_{-2.21}^{+2.92}$(-6) & 1.83$_{-0.68}^{+3.96}$(-8) & 6.47$_{-1.56}^{+2.06}$(-6) \\
    \noalign{\smallskip}
    6.72--8.40   & 3.01(4) & 9.3$_{-2}^{+1}$  & 1.09$_{-0.40}^{+0.28}$(-9)  & <5.44(-11)                  & <0.05                  & 1.54$_{-0.32}^{+0.90}$(-10) & <1.50(-9)                   & 3.16$_{-2.16}^{+6.84}$(-5) & 7.74$_{-1.59}^{+7.71}$(-8) & 2.65$_{-1.95}^{+7.41}$(-5) \\
    \noalign{\smallskip}
    8.40--10.08  & 2.23(4) & 10.4$_{-1}^{+1}$ & 9.06$_{-1.86}^{+2.35}$(-10) & <4.53(-11)                  & <0.05                  & 1.67$_{-0.34}^{+0.97}$(-10) & <9.98(-10)                  & 5.00$_{-2.49}^{+4.98}$(-5) & 1.00$_{-0.21}^{+1.00}$(-7) & 3.00$_{-2.01}^{+6.06}$(-5) \\
    \noalign{\smallskip}
    10.08--11.76 & 1.70(4) & 10.5$_{-1}^{+4}$ & 4.21$_{-0.87}^{+4.20}$(-10) & <2.11(-11)                  & <0.05                  & 8.48$_{-1.74}^{+2.19}$(-11) & <7.54(-10)                  & 5.00(-5) & 1.00$_{-0.21}^{+1.00}$(-7) & 1.06$_{-0.31}^{+0.44}$(-5) \\
    \noalign{\smallskip}
    11.76--13.44 & 1.33(4) & 11.2$_{-3}^{+6}$ & 1.30$_{-0.27}^{+1.97}$(-10) & <6.52(-12)                  & <0.05                  & 3.12$_{-0.64}^{+6.75}$(-11) & <5.65(-10)                  & 5.00(-5) & 1.00$_{-0.21}^{+1.51}$(-7) & 3.54$_{-1.03}^{+1.46}$(-6) \\
    \noalign{\smallskip}
    13.44--15.12 & 1.07(4) & 12.3$_{-6}^{+6}$ & 1.85$_{-0.93}^{+12.88}$(-11)& <9.28(-13)                  & <0.05                  & --                          & --                          & -- & 1.00$_{-0.21}^{+0.26}$(-7) & -- \\
    \noalign{\smallskip}
    \hline
    \noalign{\smallskip}
    \multicolumn{11}{c}{West side}\\
    \noalign{\smallskip}
    \hline
    \noalign{\smallskip}
    0.00--1.68 & 1.10(5) & 7.5$_{-1}^{+1}$   & 3.31$_{-0.68}^{+0.86}$(-11) & 1.13$_{-0.42}^{+0.29}$(-11) & 0.34$_{-0.14}^{+0.13}$  & 1.57$_{-0.78}^{+0.41}$(-11) & 5.00$_{-\inf}^{+74.24}$(-11) & 1.45$_{-0.76}^{+1.58}$(-7) & 2.90(-10) & 1.64$_{-0.51}^{+0.73}$(-7) \\
    \noalign{\smallskip}
    1.68--3.36 & 1.00(5) & 7.5$_{-2}^{+1}$   & 1.52$_{-0.31}^{+0.39}$(-10) & 3.16$_{-0.65}^{+0.82}$(-11) & 0.21$_{-0.06}^{+0.08}$  & 5.83$_{-1.20}^{+1.51}$(-11) & <2.52(-10)                   & 1.18$_{-0.49}^{+0.83}$(-6) & 2.37$_{-1.62}^{+7.05}$(-9) & 8.03$_{-1.94}^{+2.56}$(-7) \\
    \noalign{\smallskip}
    3.36--5.04 & 7.60(4) & 8.0$_{-2}^{+1}$   & 6.75$_{-1.39}^{+1.75}$(-10) & 3.16$_{-0.65}^{+0.82}$(-11) & 0.05$_{-0.01}^{+0.02}$  & 1.00$_{-0.21}^{+0.26}$(-10) & <3.15(-10)                   & 2.51$_{-1.80}^{+63.95}$(-5) & 5.02$_{-1.03}^{+1.30}$(-8) & 1.78$_{-1.43}^{+7.33}$(-5) \\
    \noalign{\smallskip}
    5.04--6.72 & 4.86(4) & 8.0$_{-1}^{+1}$   & 2.68$_{-0.55}^{+0.69}$(-10) & <1.26(-11)                  & <0.05                   & 1.27$_{-0.26}^{+0.33}$(-10) & <1.26(-10)                   & 5.00(-5) & 1.00$_{-0.21}^{+0.26}$(-7) & 8.66$_{-1.12}^{+1.28}$(-6) \\
    \noalign{\smallskip}
    6.72--8.40 & 2.94(4) & 9.0$_{-2}^{+2}$   & 1.64$_{-0.99}^{+0.42}$(-10) & <7.70(-12)                  & <0.05                   & 8.73$_{-1.80}^{+2.26}$(-11) & --                           & -- & 1.00$_{-0.21}^{+0.26}$(-7) & -- \\
    \noalign{\smallskip}
    8.40--10.08& 1.81(4) & 9.5$_{-2}^{+6}$   & 5.97$_{-1.23}^{+3.49}$(-11) & <2.80(-12)                  & <0.05                   & 7.02$_{-1.44}^{+1.82}$(-11) & --                           & -- & 1.00$_{-0.37}^{+0.58}$(-7) & -- \\
    \noalign{\smallskip}
    \hline
    \end{tabular}
    \tablefoot{
    \tablefoottext{a}{The deuteration ratio of \NtHp, $X$(\NtDp)/$X$(\NtHp).}
    \tablefoottext{b}{It reads 1.06$\times$10$^5$.}
    }
\label{tab:rad_model}
\end{sidewaystable*}

\end{appendix}

\end{document}